\documentclass[manuscript]{aastex}

\usepackage{rotating}
\usepackage{emulateapj5}

\slugcomment{Accepted for the Astrophysical Journal, July 10, 2013}

\shorttitle{Exoplanetary Transmission Spectroscopy}
\shortauthors{Deming et al.}

\begin{document}


\title{Infrared Transmission Spectroscopy of the Exoplanets HD\,209458b and XO-1b Using the Wide Field Camera-3 on 
the Hubble Space Telescope}


\author{Drake~Deming\altaffilmark{1,2}, Ashlee~Wilkins\altaffilmark{1}, Peter~McCullough\altaffilmark{3}, 
 Adam~Burrows\altaffilmark{4}, Jonathan~J.~Fortney\altaffilmark{5},  Eric~Agol\altaffilmark{6}, 
  Ian~Dobbs-Dixon\altaffilmark{6,7}, Nikku~Madhusudhan\altaffilmark{8}, Nicolas~Crouzet\altaffilmark{3},
Jean-Michel~Desert\altaffilmark{9,10}, Ronald~L.~Gilliland\altaffilmark{11}, Korey~Haynes\altaffilmark{12,13}, 
Heather~A.~Knutson\altaffilmark{9}, Michael~Line\altaffilmark{9}, Zazralt~Magic\altaffilmark{14,15},
  Avi~M.~Mandell\altaffilmark{13}, Sukrit~Ranjan\altaffilmark{16}, David~Charbonneau\altaffilmark{16}, 
Mark~Clampin\altaffilmark{13}, Sara~Seager\altaffilmark{17}, and Adam~P.~Showman\altaffilmark{18}}


\altaffiltext{1}{Department of Astronomy, University of Maryland, College Park, MD 20742 USA; ddeming@astro.umd.edu}
\altaffiltext{2}{NASA Astrobiology Institute's Virtual Planetary Laboratory}
\altaffiltext{3}{Space Telescope Science Institute, Baltimore, MD 21218 USA}
\altaffiltext{4}{Department of Astrophysical Sciences, Princeton University, Princeton, NJ 08544-1001 USA}
\altaffiltext{5}{Department of Astronomy and Astrophysics, University of California, Santa Cruz, CA 95064 USA}
\altaffiltext{6}{Department of Astronomy, University of Washington, Seattle, WA 98195 USA}
\altaffiltext{7}{NASA Astrobiology Institute}
\altaffiltext{8}{Yale Center for Astronomy \& Astrophysics, Yale University, New Haven, CT 06511 USA}
\altaffiltext{9}{Division of Geological and Planetary Sciences, California Institute of Technology, Pasadena, CA 91125 USA}
\altaffiltext{10}{Sagan Fellow}
\altaffiltext{11}{Center for Exoplanets and Habitable Worlds, The Pennsylvania State University, University Park, PA 16802 USA}
\altaffiltext{12}{Department of Physics and Astronomy, George Mason University, Fairfax, VA 22030 USA}
\altaffiltext{13}{NASA's Goddard Space Flight Center, Greenbelt, MD 20771 USA}
\altaffiltext{14}{Max-Planck-Institut f\"ur Astrophysik, 85741 Garching, GERMANY}
\altaffiltext{15}{Research School of Astronomy \& Astrophysics, Weston ACT 2611, AUSTRALIA}
\altaffiltext{16}{Harvard-Smithsonian Center for Astrophysics, Cambridge, MA 02138 USA}
\altaffiltext{17}{Department of Earth, Atmospheric and Planetary Sciences, Massasschusetts Institute of Technology, 
          Cambridge, MA 02139 USA}
\altaffiltext{18} {Lunar and Planetary Laboratory, University of Arizona, Tucson, AZ 85721 USA}


\begin{abstract}
  Exoplanetary transmission spectroscopy in the near-infrared using
  Hubble/NICMOS is currently ambiguous because different observational
  groups claim different results from the same data, depending on
  their analysis methodologies. Spatial scanning with Hubble/WFC3
  provides an opportunity to resolve this ambiguity.  We here report
  WFC3 spectroscopy of the giant planets HD\,209458b and XO-1b in
  transit, using spatial scanning mode for maximum photon-collecting
  efficiency.  We introduce an analysis technique that derives the
  exoplanetary transmission spectrum without the necessity of
  explicitly decorrelating instrumental effects, and achieves nearly
  photon-limited precision even at the high flux levels collected in
  spatial scan mode.  Our errors are within 6\% (XO-1) and 26\%
  (HD\,209458b) of the photon-limit at a resolving power of
  $\lambda/{\delta\lambda} \sim 70$, and are better than 0.01\% per
  spectral channel.  Both planets exhibit water absorption of
  approximately 200 ppm at the water peak near 1.38\,$\mu$m. Our
  result for XO-1b contradicts the much larger absorption derived from
  NICMOS spectroscopy. The weak water absorption we measure for
  HD\,209458b is reminiscent of the weakness of sodium absorption in
  the first transmission spectroscopy of an exoplanet atmosphere by
  Charbonneau~et~al.~(2002).  Model atmospheres having
  uniformly-distributed extra opacity of 0.012~cm$^{2}$\,g$^{-1}$
  account approximately for both our water measurement and the sodium
  absorption.  Our results for HD\,209458b support the
  picture advocated by Pont et al. (2013) in which weak molecular
  absorptions are superposed on a transmission spectrum that is
  dominated by continuous opacity due to haze and/or dust.  However,
  the extra opacity needed for HD\,209458b is grayer than for
  HD\,189733b, with a weaker Rayleigh component.
\end{abstract}



\section{Introduction}

From the first discovery of transiting extrasolar planets
\citep{charb00, henry}, transmission spectroscopy was anticipated as a
potential technique to probe their atmospheres
\citep{seager-sasselov}.  Indeed, transmission spectroscopy was used
to make the first detection of an exoplanetary atmosphere
\citep{charb02}, via optical sodium absorption observed using the
Hubble Space Telescope (HST), and sodium and potassium absorption
measurements are now possible from the ground (e.g.,
\citealp{redfield, snellen08, sing11, sing12}).  Using HST data,
\citet{barman} identified water absorption near 1\,$\mu$m in the giant
exoplanet HD\,209458b, and \citet{desert08} searched for evidence of TiO/VO
absorption in that planet.  Expanding HST transmission spectroscopy to longer
infrared (IR) wavelengths, \citet{swain08a} obtained results indicating
water and methane absorption near 2\,$\mu$m in the giant exoplanet
HD\,189733b. Similarly, \citet{tinetti} derived water and
carbon dioxide absorption near 1.4\,$\mu$m during the transit of XO-1b.

Successful transmission spectroscopy of giant exoplanetary atmospheres
is a crucial first step toward eventual spectroscopy of a nearby
habitable super-Earth using the James Webb Space Telescope (JWST,
\citealp{deming09}).  However, the IR transmission spectroscopy using
HST/NICMOS \citep{swain08a, tinetti} was challenged by
\citet{gibson11} who emphasize that the reported absorption features
are sensitive to corrections for instrumental systematic errors.
Moreover, \citet{gibson11} argue that corrections for instrumental
error cannot be made using simple linear basis models because the
`instrument model' is not sufficiently well
understood. \citet{crouzet} concluded that NICMOS instrumental
signatures remain comparable with the expected amplitude of molecular
signatures, even after a decorrelation analysis.  New methodology
\citep{gibson12a, waldmann12, waldmann13} improves the reliability of
NICMOS analyses.  However, NICMOS is no longer in operation, so
continuing the NICMOS observations {\it per se} is not possible.

Fortunately, transmission spectroscopy from 1.1 to 1.7\,$\mu$m -
largely overlapping the NICMOS G141 grism region at 1.2-1.8\,$\mu$m -
is possible using Wide Field Camera 3 (WFC3) on HST \citep{berta,
gibson12b}.  Moreover, the WFC3 detector is known to exhibit a more
uniform intra-pixel sensitivity response than does NICMOS \citep{pm1},
giving reason to expect that WFC3 observations may be less affected by
instrumental signatures.  Also, WFC3 can now be operated in a spatial
scan mode \citep{pm2} wherein the target star is trailed during each
exposure by telescope motion perpendicular to the direction of
dispersion.  Exoplanet host stars are often bright, and the spatial
scan allows the longest practical exposures for bright stars without
saturating the detector, greatly increasing the overall efficiency of
the observations.

In this paper we report WFC3 transmission spectroscopy for two
exoplanets observed in our Large (115 orbit) HST Cycle-18 program.  By
coincidence, these planets (XO-1b and HD\,209458b) were both scheduled
for observation late in our program, permitting us to acquire the
spectra in the newly-developed spatial scan mode.  One of these
planets (XO-1b) is the same as observed in transmission using NICMOS
spectroscopy \citep{tinetti, gibson11, crouzet}.  In addition to the great
photon-collecting efficiency provided by spatial scan mode, we have
achieved some new insights in the analysis of WFC3 data, beyond the
valuable methodology introduced by \citet{berta}.

We here report robust exoplanetary transmission spectra using WFC3 in
spatial scan mode.  We achieve a level of precision closely
approaching the limit imposed by photon statistics, even for these
large exposure levels collected in spatial scan mode.  Moreover, our
analysis requires no explicit decorrelation using an `instrument
model', nor does it require that the pattern of systematic errors be
consistent between orbits, or that we omit the first orbit per visit
from our analysis.

Sec.~2 describes the circumstances of our observations, and Sec.~3
discusses the initial calibration of the data, including a brief
discussion concerning the nature of instrumental signatures produced
by WFC3.  In Sec.~4 we describe new methodology that we have used to
extract the transmission spectra of the planets, and Sec.~5 gives our
results and relates them to previous work. Sec.~6 interprets our
results using model atmospheres for the planets, and Sec.~7
summarizes and comments on future possibilities.

\section{Observations}

Observations of HD\,209458b and XO-1b used WFC3 with the G141 grism,
providing wavelength coverage from 1.0 - to 1.7\,$\mu$m in first
order.  Each star was observed during a single visit comprising 5
consecutive orbits. The observational sequence in each visit began
with an undispersed image of the star using the F139M filter (central
wavelength of 1390\,nm, and FWHM = 32\,nm). The filter choice for this
image is not crucial, because its purpose is merely to establish the
position of the undispersed stellar image, used in wavelength
calibration.  Following the undispersed exposure, we obtained a
sequence of exposures using the grism, with the star scanned
perpendicular to the dispersion, under control of the HST
fine-guidance system.

The grism exposures used subarray readouts of the detector to maximize
efficiency.  Information on the number of exposures, the duration of
each sequence, subarray size, and range of planetary orbital phases
are given in Table~1.  Table~1 also includes the spatial scan
parameters and average exposure level in the spectral images, because
this information may be useful to subsequent observers.

\section{Initial Data Processing}

WFC3 grism spectroscopy samples the detector `up the ramp', i.e.,
reading each pixel non-destructively multiple times.  We process these
data by a method described in the Appendix.  After the initial
processing, the spectral frames have wavelength in one dimension, with
the spatial scan in the orthogonal direction.

Examples of the 2-D spectral frames are shown in Figure~1.  One
characteristic of spatial scan mode is that the rate of telescope
motion is not perfectly uniform, but varies slightly with time due to
jitter in the control by the fine guidance system.  Evidence of this
variation is seen on the rightmost image of Figure~1, that shows the
difference between two consecutive scans of HD\,209458b.  The variable
scan rate results in variable flux as a function of row number,
typically varying by $\pm1\%$ as shown in the plot on Figure~1.
Fortunately, because the scan is perpendicular to the direction of
dispersion, this phenomenon does not contribute significant noise to
our analysis, but it does affect how we process the data to
discriminate against hot pixel and energetic particle hits, as the
Appendix describes.

\subsection{Wavelength Calibration and Flat-fielding}

Following the initial data processing described above, we apply
wavelength and flat-field calibrations.  Each visit includes an
undispersed image of the star, and the wavelengths in the grism
spectrum are reliably fixed relative to the position of the undispersed
image.  We calculate those wavelengths using coefficients recommended
by STScI \citep{kuntschner}, with some modifications.  \citet{wilkins} found that the
original recommended coefficients did not produce optimal agreement
between the overall profile of the grism response in observed spectra
when compared to the known profile of that response, and also did not
give exact agreement with the known wavelengths of stellar absorption
lines. \citet{wilkins} varied the coefficients by up to 10\% by
trial-and-error, to achieve optimal agreement.  We have used the
\citet{wilkins} coefficients, and we obtain good consistency of the
grism response, and good accounting of stellar absorption lines (e.g.,
Paschen-beta, see below).

In addition to wavelength calibration, we apply the {\it
wavelength-dependent} flat field calibration as recommended by STScI.
Note that flat-fielding is {\it not} included in STScI pipeline
processing, even for the files of non-spatial-scan data,
because it is a function of where each target's spectrum happens to fall on
the detector, and must be done at the user stage of data analysis.  To
the extent that the stellar spectra were fixed on the detector during
each visit, and jitter and drift in wavelength were negligible,
flat-fielding would not be needed in our analysis.  In that (ideal)
case, we would not apply the flat-fielding step, since it has been our
experience that the more the data are processed, the more difficult it
becomes to achieve photon-limited results.  However, wavelength jitter
in spatial scan mode can be larger than in non-scanned observations,
so the flat-fielding step of the analysis is prudent.  However, we
have repeated the entire analysis of this paper without the
flat-fielding step, and we find consistent results in the two cases.

\subsection{Instrumental Signatures}

Our analysis is designed to be insensitive to effects caused by the
instrument and detector. To understand how we minimize such effects,
we must briefly discuss what is known about instrumental signatures in
WFC3 G141 grism data.  Figure~2 shows a portion of the transit of
WASP-18b, which was observed in our program using non-scanned mode, and
which clearly illustrates this effect (our spatial scan data show it
less clearly).  The dominant systematic error is an increase in
intensity during each group of exposures that are obtained between
buffer dumps. This pattern is shaped like a `$\Gamma$', and is
slightly reminiscent of a fish hook.  It may be physically similar to
the ramp effect seen prominently in Spitzer 8\,$\mu$m data, but the
WFC3 time scale is shorter and we cannot be certain that it has the
same physical cause as the Spitzer ramp.  We therefore use different
terminology, and refer to this WFC3 phenomenon as the `hook'.

The amplitude of the hook has been studied as a function of exposure
level by \citet{wilkins}.  They find that the hook is, on
average, zero when the exposure level per frame is less than about
30,000 electrons per pixel.  Above that exposure level, the amplitude
of the hook increases with greater exposure levels, albeit with
relatively large scatter in the relation. Our exposure levels are about 40,000
electrons per pixel (Table~1).  Based on the \citet{wilkins} results,
we expect the hook to be weakly detectable in our data.

\vspace{0.5in}
\section{Derivation of the Exoplanetary Spectra}

The principle of our analysis is that we first define and fit the
transit as observed in the integral intensity of the grism spectrum
over wavelength.  Our subsequent analysis removes this `white light'
transit, and solves for differential transit depths as a function of
wavelength.  We construct the exoplanetary transmission spectrum by
adding these differential amplitudes to the white light amplitude to
produce the transit depth versus wavelength.  There is precedent for
this method, from \citet{richardson07}. In what follows, we describe
the steps of our analysis in detail, but we also provide a concise
summary in Sec.~4.7.

Our work has benefitted from experience and analyses of non-scanned
data in our program \citep{line, mandell, ranjan, wilkins}, as well as
the analyses by \citet{berta} and \citet{gibson12b}.  Some of the
lessons learned in those analyses are: that the hook is usually
common-mode to different wavelengths and will cancel in an appropriate
ratio.  Also, derivation of the exoplanetary spectrum at the edges of
the bandpass can be problematic for two reasons.  First, the grism
spectrum exhibits jitter in wavelength that especially affects the
strongly-sloped edges of the band and must be corrected as part of the
analysis.  Second, the intensities decrease at the edges of the grism
response, so the amplitude of the hook for pixels sensing those
wavelengths will be less, and potentially not common-mode with the
remainder of the spectrum.

Following the application of wavelength and flat-field calibration to
the 2-D spectral frames, we sum each spectral frame spatially (i.e.,
along columns) to derive 1-D spectra. Mindful of the experience
related above, we perform the spatial sum over a range of rows
slightly {\it less} than the full spatial extent of the scan.  In this
manner, we avoid including lower intensities that occur at the edges
of the scan.  We sum in wavelength over a range that exceeds the
region of significant grism response, and we restrict the wavelength
range (and thus intensity range) in a later step of the analysis (see
below).  We include the first orbit of each visit in our analysis, but
as a precaution we omit the first 5 frames of the first orbits, where
the hook is strongest. 

\subsection{The White Light Transit}

Figure~3 shows the white light photometry for our observed transits of
each planet.  We have not attempted to correct for the hook, using a
divide-out-of-transit (divide-oot) procedure as described by
\citet{berta}.  The hook in these spatial scan data is weak and
inconsistent (Figure~3), and good fits are possible without making
corrections.  To fit the white light transit of HD\,209458, we
generate theoretical transit curves using the formulation of
\citet{mandel}, using non-linear limb darkening coefficients
\citep{claret}, calculated from the emergent intensities of a 3-D
hydrodynamic model of the stellar atmosphere \citep{hayek}. We
integrated the model intensities over our specific WFC3 bandpass, and
fitted the results to the nonlinear law to obtain the coefficients.
To quantify the degree to which the results for HD\,209458 depend on
the treatment of limb darkening, we also do the fits using a linear
limb-darkening law \citep{claret-bloemen}.  We find good agreement in
results between the linear and nonlinear limb darkening treatment for
HD\,209458 (see below).  Nonlinear coefficients for XO-1 would require
more interpolation in the 3-D model atmosphere grids than we prefer,
and our observed noise level is higher for XO-1 than for HD\,209458.
We therefore use linear limb darkening for XO-1.  For both HD\,209458
and XO-1, we interpolate the linear coefficients based on values from
adjacent bandpasses (J and H).

For the linear limb darkening coefficients, a range of values are
available that vary (typically by $\pm0.03$) depending on whether they
are calculated from PHOENIX versus ATLAS models, and depending on
minor parameters such as microturbulence, and on the method used to
fit the model atmosphere to the linear law. Our adopted values for
linear coefficients correspond to the middle of the range tabulated by
\citet{claret-bloemen}. A similar range of non-linear coefficients is
not available, so we use the variation of the linear coefficient to
estimate the impact of limb darkening uncertainties.

For fitting the HD\,209458b transit, we adopt the orbital parameters
($P$, $i$, $a/R_s$) from \citet{knutson07a}, with linear limb
darkening coefficient 0.28. For the non-linear case we use
coefficients calculated for our specific bandpass, as noted above.
Our WFC3 observations show a visit-long downward trend. Omitting the
first orbit, we are able to fit the remaining data using a linear
baseline plus the transit. (Note, however, that we do not omit the
first orbit from the spectrum derivation discussed below.) For XO-1b
we use the orbital parameters from \citet{burke}, with linear limb
darkening coefficient of 0.31.  We include the first orbit for XO-1.
For each planet, with the limb darkening and orbital parameters fixed,
we vary $R_p/R_s$ and the time of transit center (to account for
imprecision in the ephemeris). The resulting best-fit transit curves
are overplotted (blue lines) on Figure~3.

To estimate the errors on our derived values, we use a residual
permutation (`prayer-bead') method \citep{gillon}.  Also, we manually
vary the linear limb darkening coefficient over the range tabulated by
\citet{claret-bloemen}, and note the impact on $R_p/R_s$.  Our adopted
errors are the quadrature sum of the variation as a function of linear
limb-darkening coefficient, and the prayer-bead errors.  For
HD\,209458b we derive $R_p/R_s = 0.1209\pm0.0005$ using non-linear
limb darkening, and $R_p/R_s = 0.1214\pm0.0005$ using linear limb
darkening.  Within the errors, our results agree with other IR transit
results. \citet{crossfield} derived $R_p/R_s=0.1218\pm0.0014$ at
24\,$\mu$m from several Spitzer transits. For XO-1b, we derive
$R_p/R_s = 0.1328\pm0.0006$, that agrees (within the errors) with a
transit observed using NICMOS by \citet{burke} ($R_p/R_s =
0.1320\pm0.0005$).  Our retrieved transit times are close (tens of
seconds) to the predictions using the orbital parameters cited above.
Our results for radius ratio and transit time are summarized in
Table~2.

\subsection{Wavelength Shifts}

Upon summing the 2-D frames in the scan direction to produce 1-D
spectra, we find that these spectra are not coincident in wavelength
as judged by the displacement of the grism response curve. Variations
up to $\pm1$-column (i.e., in the wavelength direction) occur over the
span of each visit, as we measure using least-squares fitting of a
template spectrum (see below).  For HD\,209458, these variations are
almost two orders of magnitude larger than similar variations
($\pm0.02$-columns) seen in non-scanned data.  That is reasonable,
since it should be much easier for the fine guidance system to hold a
fixed position in wavelength when it does not have to scan spatially. 

A high quality analysis of spatially scanned data must account for the
wavelength shifts.  We proceed as follows.  First, we construct a
template spectrum by averaging spectra that occur before first contact
or after fourth contact, by up to one hour.  This averaging does not
account for the wavelength jitter.  Hence, the template spectrum is
slightly broadened by the jitter.  Restricting the temporal range of
the out-of-transit spectra used for the template to within one hour of
the contacts, and restricting the wavelength range to avoid the
strongly-sloping edges of the grism response, we minimize broadening
of the template due to the wavelength jitter. Broadening of the
template is not detectable in the residuals of our fits.

After forming the template spectrum we shift it in wavelength, and fit
it to all of the individual grism spectra in the visit.  For each
grism spectrum, we step through a large range of wavelength shifts of
the template, using 0.001-pixel increments, to assure that the
best-fit shift is identified.  At each shift value, the best trial fit
is achieved by a linear stretch of the template spectrum in intensity
(via linear least-squares).  No stretching of the template is applied
in the wavelength coordinate - only a shift.  The fitting process
simultaneously removes residual background intensity in the spectra
(see Appendix). The factor required for the intensity stretch
is very closely correlated to the total intensity of the system due to
the transit; the stretch factor versus orbital phase closely resembles
Figure~3.

After trying the full range of possible wavelength shifts, we pick the
best wavelength shift and linear stretch factor based on the minimum
$\chi^2$.  Shifting the template spectrum requires resampling it by
interpolation.  We use the IDL routine INTERPOLATE, with the cubic
keyword set to -0.5, this being the best approximation to ideal
interpolation using a sinc function.  Because we shift the template to
match individual spectra, re-sampling of each individual spectrum is
avoided, further minimizing the potential for adding noise via the
re-sampling process.  The wavelength shifts (in pixels) that we derive from our
fits are shown in Figure~4; the largest shifts are seen for
HD\,209458b, versus much smaller shifts for XO-1b.

Upon fitting the template spectrum to a given individual spectrum, we
subtract them and form residuals $R_{t{\lambda}}$, where $t$ indexes
time (i.e. what individual spectrum), and $\lambda$ indexes
wavelength. This subtraction removes the small amount of sky
background that survives the process described in the Appendix. This
fitting and subtraction is done separately at each time step $t$, and
it removes the wavelength variations of the grism response, as well as
canceling common-mode systematic errors (see below).  Because the flux
from the star varies with time due to the transit itself, we include a
factor to normalize the $R_{t{\lambda}}$ in units of the
out-of-transit stellar flux.  An illustration of the match between an
individual spectrum and the template is shown in Figure~5, including
the residuals in the lower panel. At each $\lambda$, the time series
$R_{t{\lambda}}$ contain the differential transits that we seek.  To
facilitate cancellation of the hook and potentially other systematic
errors, we restrict the intensity range of our analysis - hence the
wavelength range - to wavelengths whose intensity lies above the
half-power points of the grism response (see dashed lines on
Figure~5).

The fitting of the template to form the $R_{t{\lambda}}$ is key to our
analysis, because it helps to cancel common-mode systematic errors.
That cancellation is conceptually equivalent to dividing the intensity
in the grism spectrum at a given wavelength by the integral of each
grism spectrum over wavelength.  Thus, our method is similar to the
divide-oot procedure used by \citet{berta}, but is (arguably) more
general.  The divide-oot method relies on the pattern of systematic
error being consistent in {\it time}, whereas we here require that it
be common-mode in {\it wavelength}.

To see why our procedure described above is equivalent to a wavelength
ratio, consider the following.  We use a single template spectrum per
planet for each visit.  We stretch that template spectrum in intensity
for the fitting process, but the stretch is a linear factor common to
all wavelengths.  Therefore, the ratio of the template at a given
wavelength to its integral over wavelength is constant.  Moreover, the
fitting process guarantees that the wavelength integral of the
stretched template will be closely equal to the wavelength integral of
the individual spectrum being fit.  Hence, apart from a constant
factor, normalizing an individual residual by the value of the template
spectrum at that wavelength is conceptually the same as dividing the
individual spectrum at that wavelength by its integral over
wavelength.

Although we have described our method as being equivalent to a
ratio-of-wavelengths, the illumination level of the various pixels is
probably a more relevant physical variable than wavelength per se.
Since our analysis is restricted to intensities not greatly below the
peak of the grism response (Figure~5), the intensities in the data
covered by our analysis tend to be restricted to a limited range, and
this is probably the dominant factor in cancellation of systematic
errors.

Note that our analysis procedure as described above (and further
below) does not involve any explicit decorrelation versus instrument
or telescope parameters (e.g., tilt of the spectrum, Hubble orbital
phase, detector temperature, etc.).  Like the divide-oot method
\citep{berta}, we rely on cancellation of common-mode systematic
errors by operating only on the data themselves, using simple linear procedures. 

\subsection{Undersampling}

Initial correction of wavelength shifts using the above procedure
showed discordant results at some wavelengths, characterized by strong
slopes and even non-linear temporal trends in the $R_{t{\lambda}}$
values as a function of $t$.  The most discordant results occurred
near strong stellar lines such as Paschen-beta (1.28\,$\mu$m).  We
initially suspected interpolation errors in the shift-and-fit process,
but careful inspection of the uninterpolated spectra revealed that the
shapes of the stellar lines were changing as a function of the
wavelength shift, due to undersampling of the grism resolution by the
pixel grid.  The FWHM resolution of the G141 grism at 1.28\,$\mu$m
equals 2.3 pixels.  This is insufficient to eliminate changes in the
pixel-sampled line shape as a function of wavelength shift.  Figure~6
shows the Paschen-beta line in two spectra separated by about 3 hours
in the visit for HD\,209458. The change in line shape is obvious.
This line shape change is not mere noise, since it is consistent over
many spectra, and the pixel-sampled line shape changes gradually and
smoothly as a function of wavelength shift.

WFC3 sampling of 2.3 pixels per spectral resolution nominally complies
with the criterion of the Nyquist-Shannon sampling theorem
\citep{shannon}. However, the grism spectral response can violate the
premise of the theorem in the sense that its Fourier decomposition may
contain components at spatial frequencies higher than the nominal
resolution.  So the undersampling we infer here is not surprising.

Changes in the shape of pixel-sampled stellar lines during a transit
will cause noise that cannot be removed using any simple divide-oot or
ratio procedure.  Moreover, we cannot change the dimensions of the
pixel spacing.  Our solution is to force adequate sampling of the
spectral resolution, by degrading the resolution post-detection. Prior
to the analysis described above (i.e., before forming and using the
template spectrum), we convolve each 1-D spectrum with a Gaussian
kernel having FWHM = 4 pixels. We varied the width of the kernel to
determine the best compromise between supression of undersampling
errors and degradation of the spectral resolution. We apply the
convolution to the template spectrum as well as to individual
spectra. Figure~5 has been convolved with our adopted kernel.  Because
of the linearity of our analysis, we arguably could achieve similar
results by fitting the template spectrum in the presence of
undersampling errors and averaging the resultant exoplanetary spectrum
over wavelength in a subsequent step of the analysis.  We elect to
smooth the grism spectra at an early stage of the analysis because
that gives us more insight into the nature of the errors when deriving
differential transit amplitudes.

\subsection{Differential Transits}

The wavelength dependence of the transit depths is contained in the
$R_{t{\lambda}}$ residuals.  Note that the smoothing procedure
described above introduces autocorrelation in the residuals as a
function of $\lambda$ (apparent on Figure~5), but not as a function of
$t$.  At each $\lambda$, we fit a scaled transit curve to the
$R_{t{\lambda}}$, with a linear baseline.  The shape of the transit
curve is constrained to be the same as the white light (Figure~3)
transit for each planet, except that we include a correction for the
wavelength dependence of limb darkening.  We allow the amplitude of
the fitted curve to vary, since that is essential to deriving the
exoplanetary spectrum.  The best-fit amplitude and baseline slope are
found simultaneously via linear regression.  We used two different
versions of the linear baseline.  First, we used a baseline that is
linear as a function of orbital phase (called phase baseline).
Second, we used a linear baseline that is linear as a function of the
ordinal time step (called ordinal baseline).  Our derived exoplanetary
transmission spectra are insensitive to the nature of the linear
baseline (phase or ordinal), but the ordinal baseline gives about 2\%
smaller errors, so we adopt it for our final fits.  The slightly lower
errors for the ordinal baseline may indicate that the instrument
effects not cancelled by our shift-and-fit procedure depend on the
exposure number to a greater degree than they depend on mere elapsed
time.

The baseline slopes retrieved from the regression are modest, and
have little impact on our results. We also verified that the slopes are
uncorrelated with the derived exoplanetary spectral amplitudes
(Pearson correlation coefficients of about $0.15$). Note that more
sophisticated (e.g., Markov-Chain, \citealp{ford}) techniques would be
superfluous in this situation, since we are not concerned with errors
introduced by correlations between parameters, etc.  Nor do we need to
consider uncertainties on priors like the planet's orbital parameters,
because the relevant priors are already known to high precision and
are not dominant in our analysis. Instead, the dominant problem is
simply to find the best-fit differental amplitude and baseline in the
presence of noise.

Although we calculate the differential transit amplitudes by fitting a
transit curve with a linear baseline, we also checked the results
using a much simpler procedure.  Dividing the $R_{t{\lambda}}$ at each
$\lambda$ into an in-transit and out-of-transit group, we subtract the
average of the out-of-transit residuals from the average of the
in-transit residuals at each $\lambda$.  This simple in-minus-out
procedure yields transmission spectra that are very similar to the
more rigorous method of fitting a transit curve (fitting the curve
accounts for the ingress and egress portions correctly, and it permits
us to correct for the wavelength dependence of limb darkening).

There has been considerable discussion in the literature concerning
methodologies to derive exoplanetary transmission spectra (e.g.,
\citealp{gibson11, swain08a}), including some quite sophisticated
techniques \citep{gibson12a, gibson12b, waldmann12, waldmann13}.
While we respect the power of sophisticated analyses, we advocate the
virtue of making the signal visible to the eye using the simplest
linear processes.  To that end, we present Figure~7, that shows the
differential transit data for HD\,209458b, binned in intervals of 4
wavelength columns.  This is the same binning that we use for our
final spectral results.  A nominal difference is that our final
results come from fitting to single wavelengths, then averaging the
differential transit amplitudes (see below), whereas Figure~7 shows
fits to the binned data.  Since the fitting process is linear (average
of the fits equals fit to the average), there is no real difference,
and Figure~7 represents our actual results for 10 binned wavelengths
spanning the water vapor bandhead in HD\,209458b.  Notice that as
wavelength increases toward the bandhead at $\sim$\,1.38\,$\mu$m, the
differential transits change from negative (inverted) or near-zero
amplitudes, to deeper-than-average transits that are obvious by
eye. Figure~8 shows a similar comparison for XO-1b, but with more
wavelength averaging, as appropriate to the lower S/N for that planet.

As noted above, we bin the differential transit amplitudes by 4
columns to be approximately consistent with the smoothing used to
supress the detector undersampling (Sec.~4.3). The consistency is only
approximate because a square-wave binning (4 columns exactly), and a
Gaussian smoothing produce similar - but not identical - averaging.
The wings of the Gaussian kernel used in Sec.~4.3 extend beyond
$\pm2$-pixels.  Convolving also with the intrinsic 2.3-pixel
instrumental FWHM, we calculate that about about 15\% of binned
channel $N$ spills into binned channel $N+1$, and vice-versa. That
level of residual autocorrelation is not a significant factor in the
interpretation of our current results, but should be borne in mind
by future investigators using our methodology.  Our derived
exoplanetary transmission spectra for HD\,209458b and XO-1b are
tabulated in Table~3.

\subsection{Verification of Sensitivity}

Anticipating our results (Sec.~5), we find exoplanetary water
absorptions that are of significantly smaller amplitude than previous
investigators claim for the same planets.  We therefore verified the
sensitivity of our analysis by numerically injecting a synthetic
signal into our data, and we recovered it with the correct amplitude.
Specifically, we added a synthetic transit of amplitude 500 ppm,
occurring only in 10 columns of the detector spanning wavelengths
1.225-1.272\,$\mu$m, to the HD\,209458b data.  We added this synthetic
signal immediately after the stage of producing the scanned data
frames (Eq.~1 in the Appendix).  Our analysis retrieved this signal at the full
injected amplitude (not illustrated here), with the expected roll-off
at the edges of the simulated sharp band due to the smoothing used in
our analysis.  We conclude that our analysis does not numerically
attenuate exoplanetary transmission signals to any significant extent.

\subsection{Errors}

We estimate the errors associated with our differential transit
amplitudes using two techniques.  First, we calculate the standard
deviation of single points in each differential transit curve, after
the best-fit differential transit is removed (i.e., in the residuals).
We denote this value as $\sigma_1$. Then, we bin the residuals of each
transit curve over N points, varying N up to half the number of
observed points in time, and we calculate the standard deviation of
each binned set, denoted $\sigma_N$.  For Poisson errors in the
absence of red noise, we expect:

\begin{equation}
log(\sigma_N) = log(\sigma_1) - 0.5\,log(N)
\end{equation}

The slope of the observed $log(\sigma_1)$ versus $log(\sigma_N)$
relation is uncertain at a single wavelength due to the paucity of
points at high-N (i.e., few large bins).  For better statistics, we
accumulate the $\sigma$ values over all wavelengths, and show the
dependence of $\sigma_N$ versus $log(N)$ on Figure~9 for both planets.
The blue lines on Figure~9 represent the errors expected in the limit
of photon statistical noise, that decreases proportional to
$-0.5\,log(N)$. These lines are calculated {\it a priori} from the
number of electrons in the spectrum, and their close accord with the
measured scatter implies that the errors of our analysis are
close to photon-limited. Our cumulative $\sigma(N)$ values are in good
agreement with the expected photon noise and the slope of -0.5 in
Eq.(2).  One seeming difference on Figure~9 is that both planets
exhibit some $\sigma_N$ points that scatter well below the photon
noise limit at large $N$.  This occurs because the differential
transit fitting acts as a high-pass filter. Even if the differential
transit amplitude is zero, the linear regression will often find a
non-zero amplitude due to noise at large bin sizes.  The regressions
will therefore tend to remove low frequency noise as a by-product of
deriving the differential transit amplitudes.

Based on Figure~9, we calculate the error associated with each
differential transit amplitude as being the quadrature sum of the in-
and out-of-transit levels in the $R_{t{\lambda}}$ values, calculating
the errors on these levels from Eq.(2).  We also used the prayer-bead
method \citep{gillon} to calculate the error on each differential
transit amplitude.  This also indicated close agreement with the
photon limit, but the precision of the prayer-bead error estimate at a
single wavelength is limited by the relatively small number of
possible permutations.  Therefore we adopted our final errors using
the following procedure.  For each planet, we calculated the average
ratio of the prayer bead to photon errors, and the scatter about this
average.  We multiply the photon errors by this average ratio to
obtain our adopted errors for most wavelength bins.  However, a few
wavelengths exhibit prayer bead errors more than $3\sigma$ greater
than the average error level.  For these points, we use the prayer bead
error estimate for that specific wavelength.

Our derived transmission spectra for HD\,209458b and XO-1b are shown
on Figure~10, and tabulated in Table~3.  Our average error bar for the
spectrum of HD\,209458b is 36 ppm, which is 1.26 times the photon
noise.  The largest error bar for our 28 wavelength channels is 51
ppm.  For XO-1b, our average errors are 96 ppm, which is 1.06 times
the photon noise.  The largest XO-1b error bar is 111 ppm.

\subsection{Summary of Our Spectral Derivation Methodology}

To summarize our method as described above:

\begin{itemize}

\item From 2-D spectral images that are flat-fielded and
wavelength-calibrated, we make 1-D grism spectra.  We sum the 2-D
spectral images over a range slightly less than their height, to
utilize pixels having similar exposure levels, to the maximum possible
degree.  We similarly restrict our analysis to wavelengths well above
the half-intensity points on the grism sensitivity function, also to
use pixels with similar exposure levels as much as possible.

\item We integrate the grism spectra over wavelength within our
adopted wavelength range, and construct a band-integrated transit
curve.  We fit to this transit curve to obtain the white-light transit
depth ($R_p^2/R_s^2$).  We save the white-light transit depth to use below.

\item We smooth the grism spectra using a Gaussian kernel with a FWHM
= 4 pixels.  This reduces the effect of undersampling. We construct a
template spectrum from the out-of-transit smoothed spectra, and we shift it in
wavelength, and scale it linearly in intensity, to match each
individual grism spectrum, choosing the best shift and scale factors
using linear least-squares.  We subtract the shifted and scaled
template to form residuals, and normalize the residuals by dividing by
the template spectrum.  This procedure removes the white-light
transit, but preserves the wavelength variation in transit depth.

\item At each wavelength, we fit a transit curve to the residuals as a
function of time, accounting for the wavelength dependence of stellar
limb darkening.  We add the amplitude of this transit curve (a
`differential amplitude') to the depth of the white-light transit from
above.  We then co-add the results in groups of 4 wavelengths (columns on
the detector) to match the smoothing described above.  The result is
the exoplanetary transmission spectrum.

\item We determine errors using a residual-permutation method,
comparing those to errors calculated by binning the residuals over
increasing time intervals (to verify an inverse square-root
dependence), and by comparing to an {\it ab initio} estimate of the
photon noise.

\item We verify the sensitivity of the method to assure that it does
not numerically attenuate the exoplanetary spectrum.  We inject
numerically an artificial spectrum into the data at the earliest
practical stage of the analysis, and we recover it at the correct
amplitude.

\end{itemize}

\section{Results for Transmission Spectra}

\subsection{Comparison to Expectations from Spitzer}

We here illustrate our observed spectra, and immediately compare them
to models that are consistent with Spitzer emergent intensity
observations of these planets.  The strong similarity between observed
and modeled spectra reinforces our conclusion (Sec.~6.1) that we are
observing real exoplanetary absorption.  In Sec.~6, we explore
comparisons with model atmospheres in more depth.

Figure~10 overplots modeled transmission spectra for each planet in
comparison to our observed results.  The modeled spectra were
calculated by Adam Burrows \citep{burrows01, burrows10, howe} based on
combining day- and night-side model atmospheres that are consistent
with Spitzer secondary eclipse observations for these two planets
\citep{burrows07,machalek}.  The day and night-side model atmospheres
were combined by equalizing their basal pressures to join them at the
terminator of the planet.  The transmittance spectrum used for
Figure~10 represents a line of sight that passes through both the day-
and night-side models.  We fit the modeled absorptions to the data by
scaling them in amplitude, and offsetting them slightly in overall
radius, but not changing the modeled {\it shape}.  The fitted
amplitude of the HD\,209458b absorption is 0.57 of the modeled value,
and for XO-1b the fitted amplitude is 0.84 of the model.  These
factors are physically reasonable, as we discuss below, and the
correspondence between the observed and modeled shape of the
absorptions is clear.

Our analysis uses simple procedures without recourse to an instrument
model, and the smoothing we implement is motivated by an effect that
we understand physically (the undersampling).  Our errors are close to
photon-limited as verified by the inverse square-root dependence when
bining, and by the comparison with the prayer-bead errors.  We are
therefore certain that Figure~10 represents the real astrophysical
absorption spectrum, especially in the case of HD\,209458b, where the
amplitude of the absorption (200 ppm) is more than 5 times the average
error per point, and the absorption is sampled by many observed
points.

In the case of XO-1b, the detection is less visually obvious than for
HD\,209458b, but is still robust.  If we fit a flat line (the no
absorption case) to the XO-1b observations, the $\chi^2$ is 64.6 for
28 degrees of freedom.  That rejects the flat line at greater than
99\% confidence.  Moreover, the total context of the observations,
including the similarity to both the model and the HD\,209458b
observations, allows us to conclude that real astrophysical absorption
is observed in XO-1b as well as in HD\,209458b.

In Sec.~6.1 we consider whether stellar activity could contribute
significantly to our derived exoplanetary spectra, or whether true
wavelength-dependent absorption in the planetary atmospheres is
dominant, and we conclude the latter.  A more elaborated comparison
with planetary models is presented in Sec.~6.

\subsection{Comparison to Previous Observational Results}

The absorptions we detect are considerably weaker than claimed by
several previous investigations. The clearest discrepancy is for
XO-1b, as illustrated on Figure~11.  This repeats our XO-1b spectrum,
overlaid on the same plot as the results from \citet{tinetti} and
\citet{crouzet}.  The large water absorption derived by
\citet{tinetti} is inconsistent with our results: such a large signal
would be obvious even at the Figure~5 stage of our analysis.
\citet{crouzet} concluded that NICMOS instrumental signatures remain
comparable with the expected amplitude of molecular features, even
after a decorrelation analysis. \citet{crouzet} derived significantly
larger errors than \citet{tinetti}, and we therefore find less
disagreement with the \citet{crouzet} spectrum.  The discrepancy
between \citet{tinetti} and our result is either due to the
intractability of NICMOS instrument signatures, or to variability in
the exoplanetary atmosphere (i.e., clouds at the terminator).

If our difference with the \citet{tinetti} results is due to variable
clouds at the planet's terminator, it is informative to convert the
required change in absorption to an equivalent number of opaque scale
heights.  \citet{tinetti} derive an absorption depth in the spectrum
of approximately 1150~ppm (from 1.28 to 1.38\,$\mu$m), whereas we
measure only $\sim$200~ppm.  In terms of equivalent planetary radii,
the difference of 950~ppm implies:

\begin{equation}
2 {\delta}R_p/R_p = 0.00095 / (R_p^2/R_s^2)
\end{equation}

and adopting $R_p^2/R_s^2 = 0.017$ for XO-1b, we find:

\begin{equation}
{\delta}R_p = 0.0279 R_p = 2360\,km
\end{equation}

So a ring of height 2360 km would have to be opaque with clouds during
our measurement, but sufficiently clear to allow water absorption at
the time of the NICMOS observation. The scale height, $kT/{\mu}mg$ is
about 260 km, if we adopt $T=1200K$ from \citet{machalek} and use a
molecular hydrogen composition ($\mu =2.32$).  In order to attribute
our difference with \citet{tinetti} to variability of the planetary
absorption, requires variable clouds around the entire terminator of
the planet extending over 9 scale heights.  We regard this as highly
unlikely from a meteorological point of view.  So we conclude that
NICMOS exoplanet spectroscopy is unreliable when analyzed using a
standard linear basis model decorrelation (e.g., \citealp{swain08a}). In
this respect we concur with previous conclusions by \citet{gibson11}
and \citet{crouzet}.

As regards HD\,209458b, our water absorption appears inconsistent with
the results of \citet{barman}.  Barman mentions that his baseline
model, that accounts for water absorption he identified near
1\,$\mu$m, predicts a peak in $R_p$ of $1.343R_J$ at 1.4\,$\mu$m.
Estimating the continuum level at $\sim 1.315\,R_J$ from Barman's
Figure~1, we project that his baseline model would predict a
1.4\,$\mu$m absorption of about 580 ppm - about three times what we
measure. We re-visit the comparison to \citet{barman} in Sec.~6.5.

Using Spitzer transit photometry, \citet{beaulieu} found evidence for
water absorption in HD\,209458b.  While our observations do not
overlap the Spitzer bandpasses, the \citet{beaulieu} band-to-band
transit depth differences require stronger water absorption than we
observe, by several times.

\section{Interpretation Using Model Atmospheres}

We now address the degree to which our results can be affected by
stellar activity (star spots, Sec.~6.1), then we turn to the
interpretation of our results using model planetary atmospheres.  We
begin model interpretation by implementing a new fast-calculation
transmission model, and validate it (Sec.~6.2).  We also re-analyze
the HST/STIS optical transmittance data for HD\,209458b from
\citet{knutson07a} (Sec.6.3), in order to combine those data with our
WFC3 results.  We then implement our new transmittance code (Sec.6.4),
and then we compare both the HST/STIS and HST/WFC3 transmittance
spectra to calculations from other models (Sec.~6.5).

\subsection{Effect of Star Spots}

Prior to discussing our results in terms of exoplanetary atmospheric
transmission, it is necessary to demonstrate that our derived
exoplanetary spectra are unlikely to be contaminated by stellar
activity.  Although no distinct crossings of star spots are apparent
in Figure~3, two effects are possible in principle.  First, crossings
of multiple small spots might occur, and could produce a significant
cumulative effect on the transit spectra.  Second, the signatures of
uncrossed star spots - not occulted by the planet during transit -
could be `amplified' by the transit phenomenon. We first consider the
latter effect, i.e. possible amplification of uncrossed star spots.

We know from solar observations that the cool umbrae of large sunspots
exhibit water vapor absorption \citep{wallace}.  So the spectra of
solar-type stars will exhibit 1.4\,$\mu$m water absorption, albeit at
a very low level. When a planet transits without crossing spot umbrae,
it increases the relative fraction of the unocculted disk that
contains spot umbrae, hence it increases the relative depth of stellar
water absorption during transit.  It can be shown that the magnitude
of the effect masquerading as exoplanetary transit spectra is closely
approximated as $A{\delta}\epsilon$, where $A$ is the fractional
coverage of the stellar disk by umbrae, $\delta$ is the relative
absorption depth of the water band in the umbral spectrum, and
$\epsilon$ is the transit depth.

We first estimate the magnitude of $\delta$ in the above expression.
Ideally, we would measure $\delta$ from observed spectra of sunspots
at 1.4\,$\mu$m, but we do not possess such spectra due to poor
telluric transmittance.  From models of cool stars \citep{allard}, we
estimate that sunspot umbrae would exhibit about 30\% relative
absorption (line core to continuum) in the 1.4\,$\mu$m water band.  We
check that estimate from other properties of sunspots.  From umbral
spectra measured in the red-optical continuum (0.87\,$\mu$m),
\citet{penn} found that umbrae are about 0.39 as intense as the
surrounding photosphere, averaged over the solar cycle (umbrae are
slightly darker at solar maximum, about 0.35 of the
photosphere). Converting this ratio to a brightness temperature, we
obtain $T_c=4380$K for the umbral continuum. We estimate the
temperature in the water line-forming region as $T_{water}=3200$K,
from the molecular rotational temperature of the water lines
\citep{wallace, tereszchuk}. Applying those brightness temperatures to
1.4\,$\mu$m, we estimate that the umbral IR continuum intensity is about
0.52 of the photosphere, and the water band core would be about 0.21
of the photospheric intensity if it were optically thick.  That is
about a factor of two stronger than we estimated from the cool star
models \citep{allard}, which is reasonable.  We conservatively adopt
the greater (i.e., optically thick) value.  The amplitude of the water
band in the umbral spectrum is about $\delta=0.31$ (0.52-0.21) of the
photospheric intensity.

HD\,209458 and XO-1 are indicated to have average activity levels in
the compilation of \citet{knutson10}.  About half of solar-type stars
in the Kepler sample are more active than the Sun based on variations
in optical light \citep{basri}. We therefore conclude that the Sun is
a reasonable analogue for HD\,209458b and XO-1, and we calculate the
average value for coverage of the solar disk by spot umbrae.  We use
the monthly compilation of sunspot areas since 1874 given by Marshall
Space Flight
Center\footnote{http://solarscience.msfc.nasa.gov/greenwch.shtml}.
Averaging these data, we find that the fractional solar disk coverage
by sunspots (including their penumbrae) is 268 ppm on average.  At the
maximum of the solar activity cycle the typical coverage value is
about 3000 ppm, more than an order of magnitude larger.  About 32\% of
the spot area is due to umbrae, based on sunspot studies
\citep{brandt}.  Therefore the umbral disk fraction $A$ is typically
82 ppm, increasing to 960 ppm at the strongest solar maxima. Umbral
areas can be difficult to estimate, due to scattered light and limited
spatial resolution.  The seminal work of \citet{howard} found a
significantly lower (300 ppm) disk coverage at solar maximum based on
Mt.~Wilson photographic plates (1921-1982), but about the same average
coverage that we infer here.

For HD\,209458b, we have $\epsilon=0.0146$ \citep{knutson07a}, so
$A{\delta}\epsilon$ is 0.4 to 4.3 ppm, and the numbers for XO-1 are
similarly small.  Even if these stars are several times more active
than the Sun at solar maximum, the amplification-of-star-spots effect
is not a major contributor to our measured water band depths.  

We now turn to the possible cumulative effect of crossing small star
spots during transit.  Occultation of water absorption in star spots
would produce an apparent {\it emission} in the derived exoplanetary
spectrum, or it would weaken real exoplanetary absorption.  The worst
case effect would occur for a star at the peak of its activity cycle,
when all of the spots on the stellar disk were occulted by the planet
during the specific phases of our WFC3 data (Figure~3). The star spots
would have to be distributed so that they were all crossed during our
partial coverage of the transit, and their sizes would have to be
sufficiently equal to prevent a single large spot-crossing being
visible.  In this unlikely case, the planetary absorption could be
weakened by 0.31(960)= 298 ppm, a significant effect.  However, if 960
ppm of star spot umbrae were occulted during transit, then the
white-light transit in Figure~3 would be more shallow because the
umbral continuum is fainter than the photosphere.  That effect would
be approximately 960(0.52)= 500 ppm in transit depth.  It would
decrease our value of $R_p/R_s$ by approximately 0.002 for both
HD\,209458b and XO-1b.  Those decreases are unlikely, given our
agreement with Spitzer IR transits of HD\,209458b \citep{crossfield},
and with an independent transit of XO-1b analyzed by \citet{burke}
(Sec.~4.1). Moreover, if the weak absorptions we derive for these
exoplanets are due to masking by star spots of intrinsically stronger
exoplanetary absorption, then the specific and unlikely
spot-occultation circumstances described above would have to apply
independently to both stars.

We conclude that neither occulted nor unocculted star spots have
significantly affected our results, and we are measuring real exoplanetary
water in transmission.

\subsection{Validation of a Spectral Transmittance Code}

In Sec.~6.3, we interpret our results using a new spectral
transmittance code. This code is intended for rapid line-by-line calculation of
transmittance spectra, so as to explore parameter space when varying
mixing ratios, cloud heights, atmospheric temperature, etc.  We intend
to use it for future WFC3 investigations as well as to illuminate the
present results.  Here, we describe the code, and the tests we have
conducted to validate it.

Our transmittance code is based on the work of \citet{richardson03};
consequently many of the algorithms it uses were tested previously.
We have modified the original code to use the slant-path geometry
appropriate for transmittance spectra.  The code uses a layered
hydrostatic equilibrium atmosphere, with pressures spaced equally in
the log from $1$ to $10^7$ dynes cm$^{-2}$ (adjustable).  It
incorporates continuous opacity due to collision-induced absorption by
H$_2$ \citep{zheng, borysow}, and it includes a provision for
calculating number densities of major molecules in thermal
equilibrium. Nevertheless, we here specify a depth-independent mixing
ratio of water in an ad-hoc fashion for the purpose of exploring
parameter space.  We include opacity for water using the line-by-line
data from \citet{part_schwenke}.  To speed the calculations, we sum
and pre-tabulate the strengths of the extremely numerous water lines
within wavelength bins, and represent each bin by a single line having
the strength of the total.  We choose a bin width of 0.1 cm$^{-1}$ ($2
\times 10^{-5}\,\mu$m at 1.4\,$\mu$m), much smaller than our WFC3
resolution. The average within each bin is represented by a Voigt
profile with a damping coefficient of 0.1 cm$^{-1}$ per
atmosphere. This averaging is valid only at a single temperature,
because each bin in wavelength will contain lines of different lower
state excitation, which adds a non-linear factor.  We thus
pre-tabulate a separate line list for each isothermal model.  From the
line and continuous opacity, we numerically integrate along the
un-refracted path that passes tangent to each layer, to calculate the
optical depth $\tau$ and extinction $exp(-\tau$) over that path.  We
adopt a source function of zero, i.e., we neglect self-emission by the
planetary atmosphere. From the optical depths, we calculate
the effective blocking area of the planet as a function of wavelength.

To validate this code, we conducted two tests.  First, we specified an
ad-hoc continuous opacity proportional to $\lambda^{-4}$, and we
conducted a test described by \citet{shabram}.  This test involves
verifying the analytic relation $dR_p/d\ln{\lambda} = -4H$, where
$R_p$ is the calculated wavelength-dependent transit radius, and $H$
is the pressure scale height. Using a 1000-layer isothermal atmosphere
with a constant scale height, we calculated the slope of $R_p$ versus
$\ln{\lambda}$, and we find agreement with the slope of -4 to
within 0.58\%.  For atmospheres having 500 and 200 layers, our
precision on the slope is 1.28 and 3.42\%, respectively.  Our grid of spectral
calculations to interpret our WFC3 spectra (Sec.~6.3) uses the
200-layer version, which is more than adequate for the present
purpose, as our second test now establishes.

For the second test, we compared our calculated water transmission
spectrum to an independent calculation from J.~J. Fortney.  The
Fortney model is new, but uses the methods described
in \citet{fortney10}.  This comparison used our 200-layer version of
the code, and an isothermal model having $T=1500$K, a surface gravity
of 10~meters~sec$^{-2}$, and a water mixing ratio (independent of
pressure) of 0.00045. The Fortney calculation uses different
algorithms, different layering, and different numerical approximations
to represent the same atmosphere. Results from the two calculations
are in excellent agreement, and shown in Figure~12, where the
(smoothed) results from the new code track the Fortney calculation
very closely.  As an additional test, we repeated the Figure~12
comparison using a calculation by N.~Madhusudhan, adopting different
atmospheric parameters, and we again achieved excellent agreement (not
illustrated here).  We therefore conclude that our new transmittance
code is validated for the purpose of this paper.

\subsection{Re-analysis of the HST/STIS Photometry}

We want to utilize our results together with transmission spectra
derived by \citet{knutson07a} from HST/STIS data.  \citet{knutson07a}
give errors for $R_p$ and $R_s$ separately, but a large fraction of
those errors are common-mode between $R_p$ and $R_s$.  To clarify the
error on $R_p/R_s$ from the STIS data, we were motivated to re-analyze
that photometry (from Table~1 of \citealp{knutson07a}).  Our re-analysis
adopts many of the \citet{knutson07a} parameters without attempting to
vary them.  Specifically, we adopt their transit center times, their
non-linear limb darkening coefficients, and their band-to-band
differences in $R_p/R_s$.  We fit their photometry for all 10 bands
simultaneously, using a Markov Chain Monte Carlo (MCMC) method with
Gibbs sampling \citep{ford}, to determine the band-averaged value of
$R_p/R_s$, as well as $a/R_s$ and $i$.  We initialize our MCMC chains
using the parameters from \citet{knutson07a}, and these good starting
values allow our chains to converge rapidly (in approximately 1200
steps). However, the chains run slowly, because we are fitting to 10
bands simultaneously using nonlinear limb darkening.  We therefore run
4 independent chains simultaneously, each having 60,000 steps, and we
combine their posterior distributions.  The combined distributions are
closely approximated by Gaussians, and yield $R_p/R_s =
0.1210\pm0.0001$.  The small error in $R_p/R_s$ reflects the large
volume of data over 10 wavelength bands.

Our re-determined value for the band-averaged $R_p/R_s$ is in mild
disagreement with \citet{knutson07a}, in the sense that our result
favors a larger $R_p$ by 1.5 times their error for $R_p$. However, our
re-determined STIS value for $R_p/R_s$ is in good agreement with
our WFC3 value, so our analysis is self-consistent.  As noted in
Sec.~4.1, we also agree with other IR determinations of $R_p/R_s$
\citep{burke, crossfield}, except for the values from
\citet{beaulieu}.

\subsection{Interpretation of HD\,209458b Using Model Atmospheres}

We now turn to exploring what our new observations imply about the
atmospheres of these planets.  In the case of XO-1b, the scaling
factor required to fit the Spitzer-based model to the data on Figure~10
was 0.84, versus 0.57 for HD\,209458b.  These factors do not per se
indicate discrepancies with the conclusions from the Spitzer
investigations \citep{knutson08,machalek}, because molecular
absorption during transit is much more sensitive to conditions such as
clouds and haze.  Given that the scale factor for XO-1b is close to
unity, and considering the larger noise level of that spectrum
compared to HD\,209458b, we conclude that XO-1b is adequately
described by extant models, not counting the \citet{tinetti}
calculations.  In contrast, the relatively small scale factor required
for HD\,209458b, combined with the lower noise level of those data,
motivate us to inquire further what our results imply for the
atmosphere of that planet.

In principle, it would be possible to analyze our water transmission
spectrum simultaneously with sodium absorption measurements
\citep{charb02, sing08, snellen08}, CO absorption data
\citep{snellen10} and Spitzer secondary eclipse photometry
\citep{knutson08} and spectroscopy \citep{richardson07, swain08b}. Such
an analysis could incorporate guidance from hydrodynamic models
\citep{showman08, dobbs-dixon, showman13, rauscher13} to account for the
longitudinal transfer of stellar irradiance, and could explore the
full parameter space of composition and temperature structure
\citep{madhu09, madhu10}, as well as the effect of clouds and hazes at
the terminator \citep{parmentier}.  However, we do not attempt such an
ambitious investigation here.  Instead, we compare the combination of
HST/STIS and HST/WFC3 measurements to theoretical transit spectra from
various models without much fine-tuning, and we discuss the discrepancies so
as to guide the general direction of more exhaustive atmospheric
modeling in the future.

The weakness of the water band we observe is reminiscent of the first
detection of this (indeed, of any) exoplanetary atmosphere by
\citet{charb02} who observed the sodium `D' line doublet near
0.5893\,$\mu$m using HST/STIS.  To account for the weakness of the
sodium feature, \citet{charb02} varied the height of an opaque cloud
layer, using the model described by \citet{brown}.  We here perform a
similar calculation, using our spectral transmittance code described
in Sec.~6.2. We vary both the pressure level for the top of an opaque
cloud layer, and the mixing ratio of water vapor, and we measure the
strength of the 1.4\,$\mu$m band at about 300 points over a 2-D grid,
calculating a full transmittance spectrum (as per Figure~12) at each
grid point. The question of the temperature structure at the
terminator of the planet - needed for this calculation - is
problematic, in spite of significant work on this topic
\citep{sing08}. Our results are not very sensitive to the T(P)
relation, so we prefer to use a simple isothermal atmosphere, based on
the observed brightness temperatures from Spitzer \citep{crossfield}.
Ideally, we would average the day-side and night-side brightness
temperatures from Spitzer to arrive at an estimate of the temperature
at the terminator.  Although Spitzer around-the-orbit observations for
HD\,209458b have been obtained, they are still under analysis.
Therefore we adopt the same relative day-to-night change as for
HD\,189733b \citep{knutson07b}, and apply that relative variation to
the HD\,209458b day side temperature from \citet{crossfield}.  This
yields an estimate of 1200K for the terminator, and Figure~13 shows a
grid of band amplitudes calculated at that temperature.

The blue line with $\pm1\sigma$ error limits on Figure~13 is the
amplitude of the observed band.  Since the models have fine-scale
structure with wavelength, and the observations have point-to-point
noise, we define a band amplitude for Figure~13 by averaging over two
wavelength regions having low and high water vapor opacity,
respectively, i.e., 1.27-1.30 and 1.36-1.44\,$\mu$m.  This results in
an observed band amplitude of 176$\pm$25 ppm.  Note that the amplitude
of the absorption near the actual bandhead at 1.38\,$\mu$m is slightly
higher, about 200 ppm.

Contours of constant mixing ratio are included on Figure~13.  The
solar-abundance contour (drawn in red) intersects the observed band
amplitude only where the cloud top pressure is low, near 1.5~mbars.
That is similar to, but less extreme, than the sodium case;
\citet{charb02} remarked that if the sodium weakness is attributed
solely to clouds, then it `would require...clouds tops above 0.4 mbar.'
\citet{fortney03} concluded that silicate and iron clouds could reside
at pressures of several millibars in the atmosphere of HD\,209458b. As
concerns alternate explanations, note that Figure~13 implies a
semi-forbidden region, where no contours pass into the lower right of
the plot. The cloud-top pressure levels depicted on the right of
Figure~13 imply a clear atmosphere, and even our lowest modeled mixing ratio
(-5.2 in the log) is not sufficient to weaken the band to account for
our observations if the atmosphere is clear. The total column density
along the line of sight at high pressures in the tangent geometry is
so large that even unrealistically small mixing ratios are
insufficient to weaken the band to the observed degree, in the absence
of other water-destruction mechanisms such as photolysis (unlikely in
the deep atmosphere).  We conclude that we are not observing a clear
atmosphere.

Figure~14 shows our full HD\,209458b transit depth spectrum
($R_p^2/R_s^2$), combining both our WFC3 and re-analyzed STIS results.
The combination of these data span wavelengths from 0.2 to 1.6\,$\mu$m
with a consistent observed lower envelope and overall level in
$R_p^2/R_s^2$.  We first add the caveat that systematic differences
might still remain between the overall level of the STIS and WFC3
transit depths, in spite of the seeming consistency.  Nevertheless,
Figure~14 represents the best composite optical/near-IR transmission
spectrum of HD\,209458b to date, so we proceed to ask what it reveals
about the exoplanetary atmosphere.

Now, we compare this combined spectrum to two models.  First, we
used a grid of spectra by Adam Burrows, based on a 1200K isothermal
temperature structure, as used above.  The grid utilizes the
methodologies described by \citet{burrows01} and \citet{burrows10} and
\citet{howe}, but it incorporates different amounts of extra gray
opacity.  We interpolate in this grid to find that an extra opacity of
0.012\,cm$^2$\,g$^{-1}$ matches the 1.4\,$\mu$m water absorption at
the bandhead, and provides suitably low absorption at
1.15\,$\mu$m. The lowering of the 1.15\,$\mu$m absorption occurs because
that intrinsically weaker band requires a longer path length to
produce significant absorption, and long path lengths are masked by
the extra opacity. A Burrows isothermal model having no extra opacity
(not illustrated) shows a much more prominent peak at 1.15\,$\mu$m. 

The profile of the 1.4\,$\mu$m band is not matched optimally by the
isothermal models, not as well as on Figure~10 for example.  The real
absorption line of sight passes through different temperatures on day
and night hemispheres of the planet. Figure~10 accounts (crudely) for
different temperatures along the line of sight, not included in the
isothermal model for Figure~14. Including that line-of-sight
temperature variation may be essential to matching the band profile.

The Burrows model on Figure~14 is sufficiently high-resolution in
wavelength to permit meaningful comparison with the sodium absorption
measured by \citet{charb02} and \citet{snellen08}.  We plot the `narrow'
band absorption from \citet{charb02} on Figure~14 (triangle point with
error bar). The \citet{snellen08} results (not plotted) are consistent
with \citet{charb02}, considering the different bandpasses.
Integrating the Burrows model having 0.012\,cm$^2$\,g$^{-1}$ extra
opacity over the band used by \citet{charb02} (red diamond point)
produces agreement within the error bar.

One aspect of the observations that are not reproduced by the simple
isothermal Burrows model with gray opacity, is the tendency toward
increasing radius in the blue and UV, at the left edge of
Figure~14. An increase of transit radius at short wavelength may be
related to the absorber that causes a temperature inversion in this
planet \citep{burrows07}. It may also be produced by Rayleigh
scattering from a population of small particles, that we do not
include in the Burrows calculations for Figure~14 (but Rayleigh
scattering by molecules is included). In this regard, we overplot a
model from Ian~Dobbs-Dixon (blue line on Figure~14), based on the
methods described in \citet{dobbs-dixon}.  This model uses a full
radiative hydrodynamic treatment of the temperature structure, which
may explain why it produces a better (but not perfect) account of the
1.4\,$\mu$m band profile. It has no extra gray opacity, but it
incorporates extra opacity of 0.004\,cm$^2$\,g$^{-1}$ at 0.8\,$\mu$m,
with a $\lambda^{-4}$ dependence.  Because that Rayleigh opacity is
concentrated at short wavelengths, we must scale the modulation in the
modeled spectrum downward by a factor of 3 to match the observed
1.4\,$\mu$m band.  That scaling is unphysical, but it allows us to
judge the relative importance of gray versus Rayleigh opacity that
will be needed to match the observations.  After scaling, the blue
line produces relatively good agreement with the 1.4\,$\mu$m data, but
overestimates the 1.15\,$\mu$m feature as well as slightly
overestimating the increase in absorption in the blue and UV.

Finally, we point out one notable discrepancy in the model
comparisons. The STIS point near 0.95\,$\mu$m cannot be reproduced
by models while still being consistent with our 1.4\,$\mu$m band measurement.
\citet{barman} argued for water vapor in HD\,209458b based in part on the STIS
data near 0.95\,$\mu$m, but \citet{knutson07a} did not claim water
detection from those same data.

From the above comparisons, we conclude that:

\begin{itemize}

\item A uniformly distributed extra opacity, gray in wavelength
  dependence, of approximate magnitude 0.012\,cm$^2$\,g$^{-1}$ is
  needed in isothermal models at 1200K in order to match our observed
  water transit absorption for HD\,209458b at 1.4\,$\mu$m, and to
  produce the weakness of water at 1.15\,$\mu$m, and account for weak
  sodium absorption in the optical.

\item Models that include realistic temperature distributions along
the line of sight may be required to match the band profiles of the
water absorption.

\item Our results are consistent with the situation described by
\citet{pont13} for HD\,189733b, wherein weak molecular absorptions are
superposed on a spectrum whose broad variations with wavelength in
transit are dominated by haze and/or dust opacity.  However, the extra
opacity required for HD\,209458b needs to be grayer than the strong
Rayleigh component needed for HD\,189733b.

\end{itemize}

\section{Summary and Further Implications}

We have demonstrated that WFC3 spectroscopy using the new spatial scan
mode can yield exoplanetary spectra whose error level falls
significantly below the 0.01\% level.  We detect 1.4\,$\mu$m water in
both HD\,209458b and XO-1b at a relatively low level of absorption,
only 200 ppm at the 1.38\,$\mu$m bandhead.  Our results for XO-1b
contradict the much larger absorption derived by Tinetti et al.(2010),
and we concur with \citet{gibson11} that NICMOS spectroscopy is
unreliable when analyzed using the standard linear basis model
approach \citep{swain08a}. \citet{fortney05} predicted that giant
exoplanetary atmospheres would contain condensates and hazes that
would `lead to weaker than expected or undetected absorption
features'.  For HD\,209458b, an atmosphere with
0.012\,cm$^2$\,g$^{-1}$ of extra opacity, gray in character, is
required in order to match the subtle water absorption we detect and
the sodium absorption in the optical. \citet{pont13} have argued that
molecular absorptions in HD\,189733b are weakened by haze and/or dust
opacity, and our results suggest a similar situation for
HD\,209458b. One difference is that haze and/or dust opacity is grayer
for HD\,209458b, with a weaker Rayleigh component as compared to
HD\,189733b.

The capability to derive transmittance spectra for planets transiting
bright stars, at an error level below 0.01\% in a single HST visit,
opens new scientific horizons.  For example, it becomes feasible to
monitor meteorological variability of conditions at the terminator of
the planet.  The transmittance spectra of super-Earths should be
detectable using multiple visits \citep{bean10, bean11, kreideberg}, even for
high molecular weight atmospheres.  Applying the spatial scan to
secondary eclipses, it should be possible to confirm the existence of
atmospheric temperature inversions via low-noise detection of the
water band profile in {\it emission} rather than in absorption.
Finally, we note that there is much discussion in the community
concerning dedicated space missions to characterize exoplanetary
atmospheres using transits.  The design of such missions should
prudently consider that molecular absorptions may be considerably
weaker than are modeled using clear atmospheres.

\section{Acknowledgements}

We are grateful for the excellent support of the staff at STScI for
the scheduling and execution of our demanding obsrvations, especially
our contact person Shelly Meyett.  We also thank an anonymous referee for
insightful comments that improved this paper.

\clearpage

\appendix{PRODUCTION OF 2-D SPECTRAL FRAMES} \\ 

We here describe how the spectral frames shown in Figure~1 are
constructed from the sample-up-the ramp data frames, and how
discrepant pixels and energetic particle hits are corrected. Each
non-destructive sample of the detector is provided to observers as an
extension in a FITS file (the *ima.fits files available from the
Multi-Mission Archive at Space Telescope, MAST).  Normally the
*ima.fits files are processed by an analysis pipeline at the Space
Telescope Science Institute (STScI), which fits the slope and makes
the resulting intensities available in *flt.fits files.  However, in
the case where the source is rapidly moving, as it is for the spatial
scan mode, the normal analysis pipeline is inapplicable because a
given pixel is not always viewing the same celestial scene, and that
invalidates the pipeline's fit to the slope at each pixel.  In scan
mode, there are two alternative methods that can be used for initial
data processing.  One method is to subtract the first read of the
detector from the last read, thus producing an image of the total
accumulated electrons on the detector during the spatial scan.
However, we use an alternate method that offers several advantages
over a simple `last minus first' subtraction.

For a given exposure, let $M$ be the number of times that each pixel
is sampled, and let $i = 1, M$ index the individual samples.  For each
$i \leq M-1$, we subtract sample $i$ from sample $i+1$, thus forming
differences $D_i = C_{i+1}-C_i$ where $C$ denotes the charge on a
given pixel in electrons.  Because scanning occurs parallel to
columns, these differences reveal the spectrum of the star over a
limited range of rows on the detector.  Considering the difference
$D_i$, we zero the rows not containing the target star, using a
top-hat mask whose width extends to the wings of the stellar point
spread function (PSF) - typically 15 pixels (1 pixel =0.14 arcsec).
We then form the sum:

\begin{equation}
S_j = \sum\limits_{i=1}^{M-1} D_i,
\end{equation} 

where $j=1,N$ indexes the $N$ times that the scan is repeated over the
duration of the transit.  Each sum is the image of the spatial scan
that we use in further analysis.

The above methodology reduces the effect of sky background.  Because
of the masking applied to each of the $D_i$, sky background not
immediately adjacent to or underlying the star is zero-ed, and does
not accumulate in the summation in Eq. (1).  Only the background near
the star in each $D_i$ survives this procedure.  However, the residual
background flux that accumulates in the brief interval between
consecutive reads of the detector is small ($<0.1$\%) compared to the
fluxes from these bright stars, and is not removed at this stage of
the analysis. It is removed by our wavelength shift correction procedure,
described in Sec.~4.2.  Removal of the residual background by that
procedure produces a small bias on the stretch factor (see Sec.~4.2), but
we verified that the bias has negligible effect (by several orders of
magnitude) on the final exoplanetary transmission spectrum. The
residual background is not removed when fitting to the white-light
transit (Sec.~4.1).  In that case it biases the white light transit
depths, but the magnitude of the bias is much less than the
observational error ($< 0.2\sigma$). Advantages of the above procedure
are that it minimizes the effect of hot pixels and energetic particle
events that would otherwise overlap the scan.  Moreover, it allows for
discrimination against other stars that are spatially resolved, but
would overlap in a simple last-minus first difference.

Although the procedure described above minimizes the effects of
discrepant pixels, it does not completely eliminate them.  Normally,
hot and transient pixels are identified and corrected via a numerical
median filter applied to the time history of each pixel.  The spatial
scan rate variations complicate that procedure because they contribute
to the intensity fluctuations of every pixel versus time, and could
interfere with the median filter process.  We therefore apply a
5-point median filter to the {\it ratio} of a given pixel intensity to
the total intensity in that row of the detector.  This ratio cancels
the scan rate variations (Figure~1), and isolates the behavior of the pixel
itself.  We apply the median filter in a two-pass process.  The first
pass corrects pixels that are discrepant by more than $10\sigma$ from
the median value, where $\sigma$ is the standard deviation of the
difference between the time history of the pixel and the
median-filtered version of that time history. The first $10\sigma$
pass serves to eliminate very large fluctuations that might perturb
the calculation of $\sigma$. The second pass uses a lower threshold,
correcting pixels that are discrepant by more than $3\sigma$. For
HD\,209458b, 0.15\% of the pixels are corrected by this procedure, and
0.04\% in the case of {XO-1}.



\renewcommand{\baselinestretch}{1.5}

\begin{sidewaysfigure}
\centering
\epsscale{0.6}
\vspace{-3.5in}
\hspace{-2.5in}
{\includegraphics[angle=90]{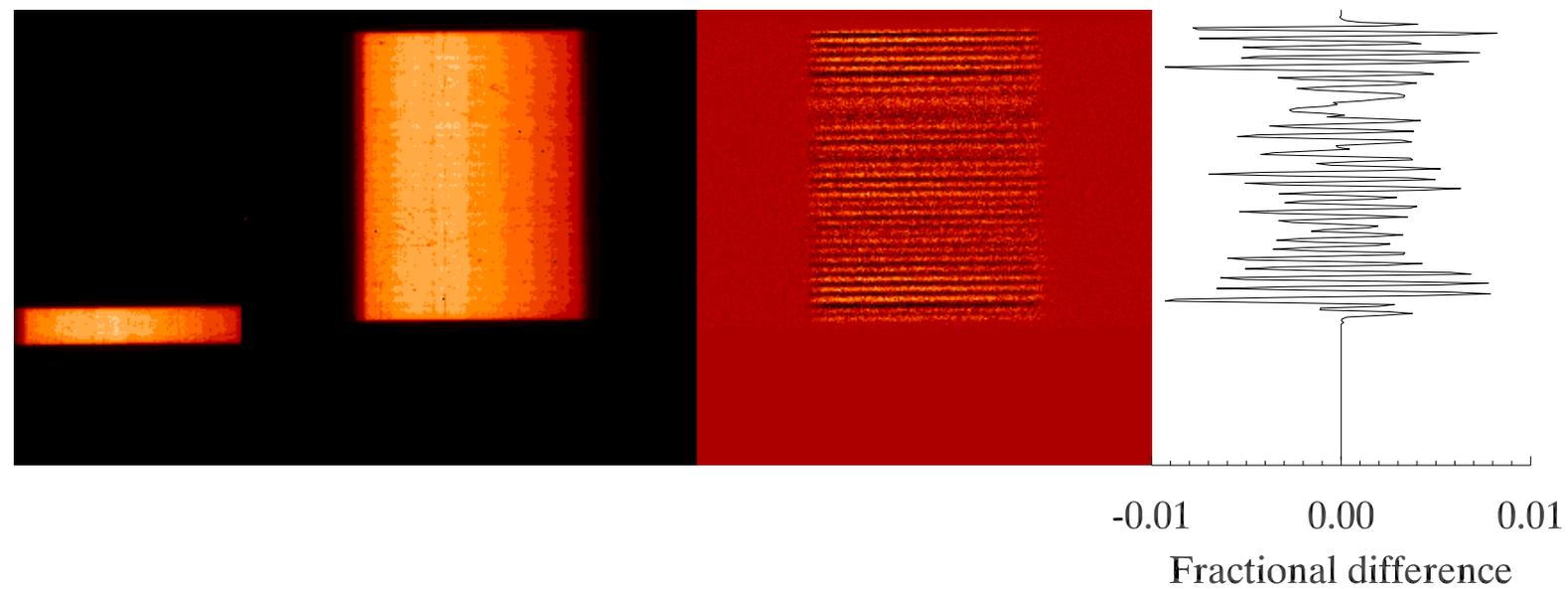}}
\vspace{1.0in}
\caption{Images of the spatial scans at representative times.  Left to
right, the images are: a scan of XO-1b (20 pixels scan height), a scan
of HD\,209458b (228 pixels scan height), and the difference between
two consecutive scans of HD\,209458b.  The difference images shows a
striped appearance due to small variations in the scan rate under
control of the Hubble fine guidance system.  The plot shows the
fractional intensity fluctuations as a function of scan position
(Y-axis) for the difference image.}
\label{fig1}
\end{sidewaysfigure}

\clearpage

\begin{figure}
\epsscale{.80}
\plotone{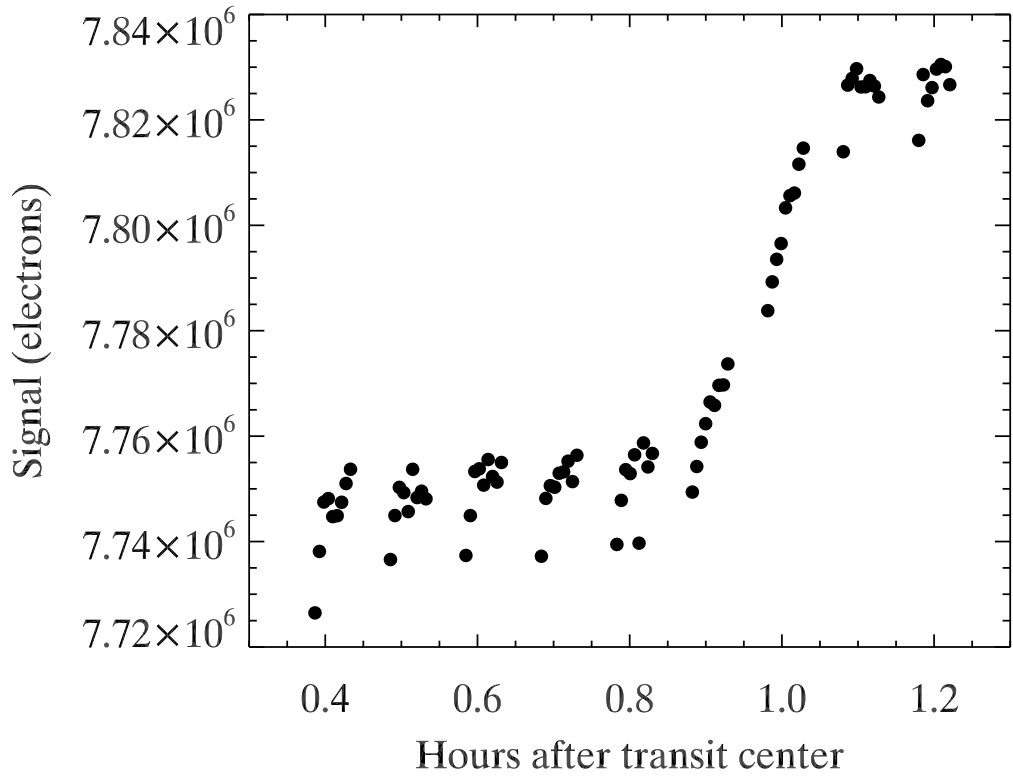}
\caption{Example of the most prominent instrument-related systematic
effect in staring-mode data from our Cycle-18 WFC3 program.  This example zooms-in
on the egress portion of a transit of WASP-18b, using the integral
intensity over each grism spectrum.  The space between groups of
observed points is due to the time needed to transfer the data buffer.
Within each group of points, the intensity increases in a pattern
shaped somewhat like `$\Gamma$'.  We call this pattern the `hook'.
\label{fig2}}
\end{figure}

\clearpage

\begin{figure}
\epsscale{.50}
\plotone{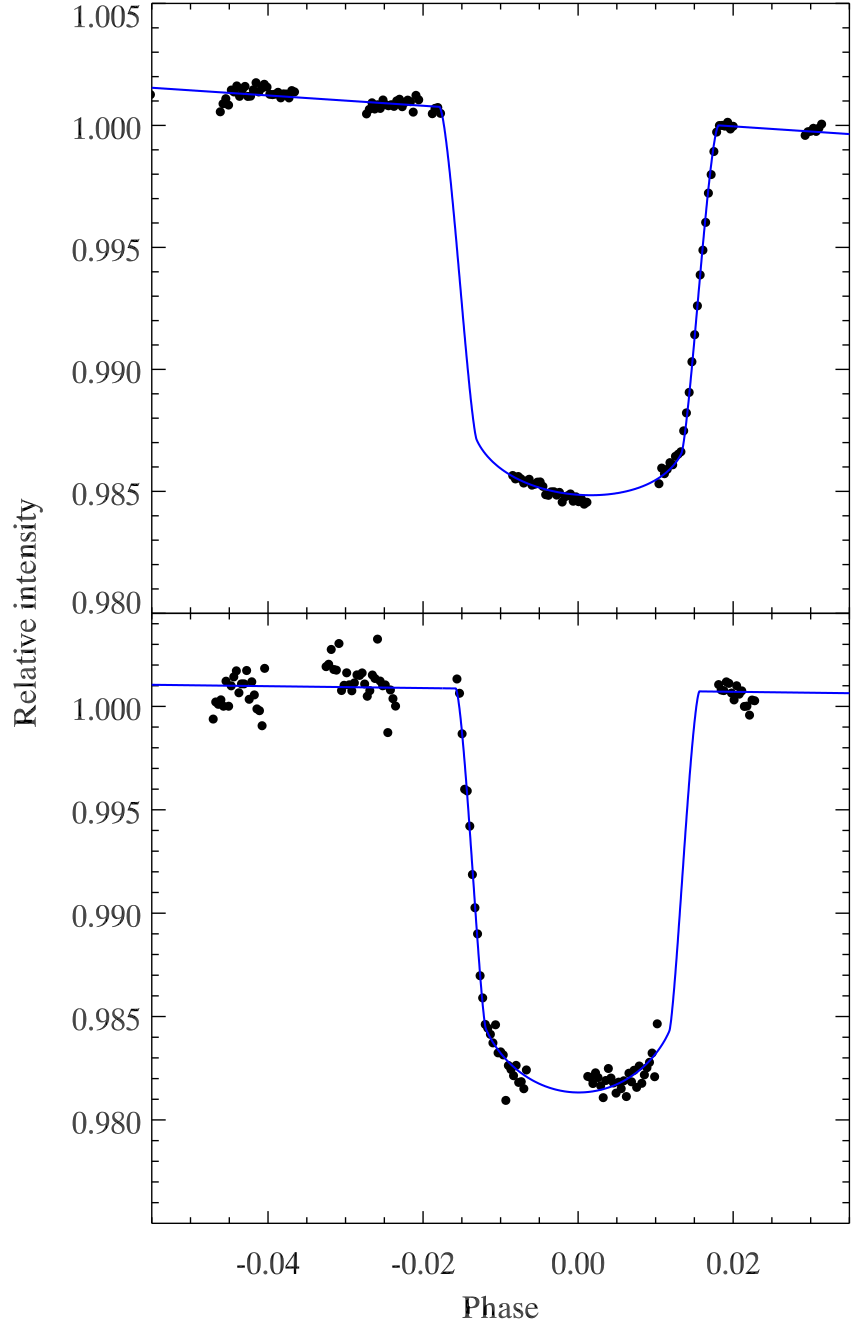}
\vspace{0.8in}
\caption{Transits observed in spatial scan mode: HD\,209458b (top) and XO-1b (bottom),
integrating over the entire grism bandpass (`white light').  No attempt was made to remove
systematic effects by divide-oot methodology \citep{berta}; these are purely `as observed'.  The
blue curves are fit to the data by varying $R_p/R_s$, and a correction to the time of center transit, 
but fixing other parameters at the values given by \citet{knutson07a} and \citet{burke}. 
\label{fig3}}
\end{figure}

\clearpage

\begin{figure}
\epsscale{.80}
\plotone{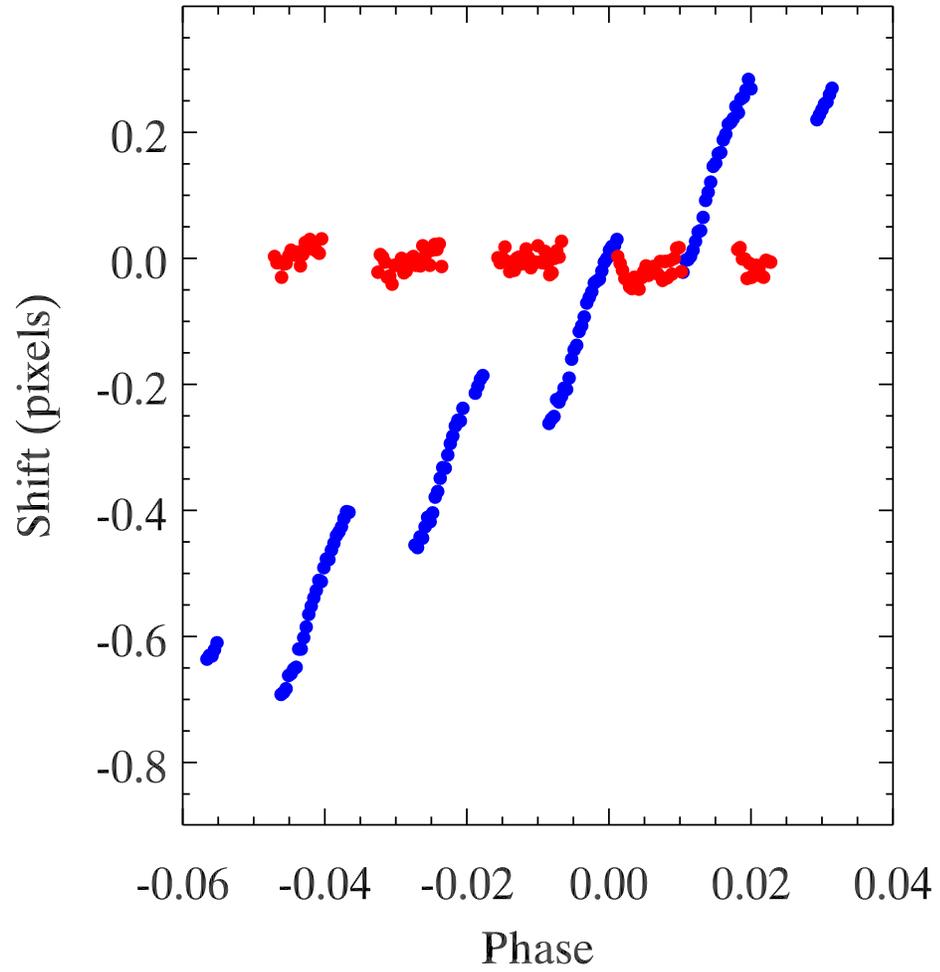}
\vspace{0.5in}
\caption{Wavelength shifts derived from our shift-and-fit procedure, 
versus orbital phase.  The grouping of the points shows the different orbits in each visit.  
Blue points are HD\,209458 and red points are XO-1.
\label{fig4}}
\end{figure}

\clearpage

\begin{figure}
\epsscale{.50}
\vspace{-0.5in}
\plotone{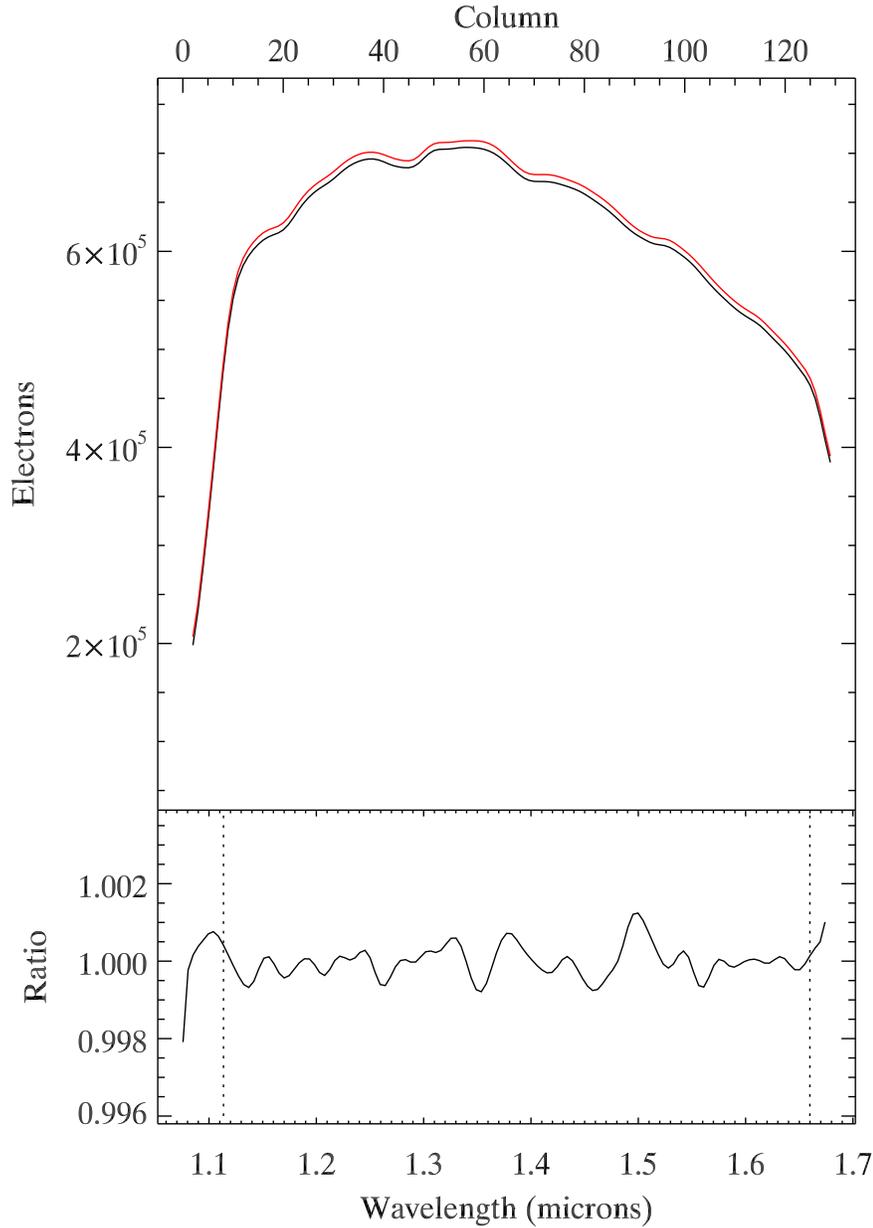}
\vspace{0.7in}
\caption{ {\it Top panel:} example of the grism spectrum of XO-1.  The
black line is a spectrum at a single time, showing the roll-off in
grism response shortward of 1.1\,$\mu$m and longward of 1.65\,$\mu$m.
The red curve is the out-of-transit template spectrum (see text),
shifted upward by 1\% for clarity of illustration. These spectra have
been smoothed using a Gaussian kernel having FWHM = 4 pixels. {\it
Lower panel:} difference between the single smoothed spectrum and best-fit
template, normalized by the intensity in the template spectrum.  The
vertical dashed lines define the wavelength range used in the analysis
of XO-1b. (A similar range was used for HD\,209458b.)
\label{fig5}}
\end{figure}

\clearpage

\begin{figure}
\epsscale{.80}
\plotone{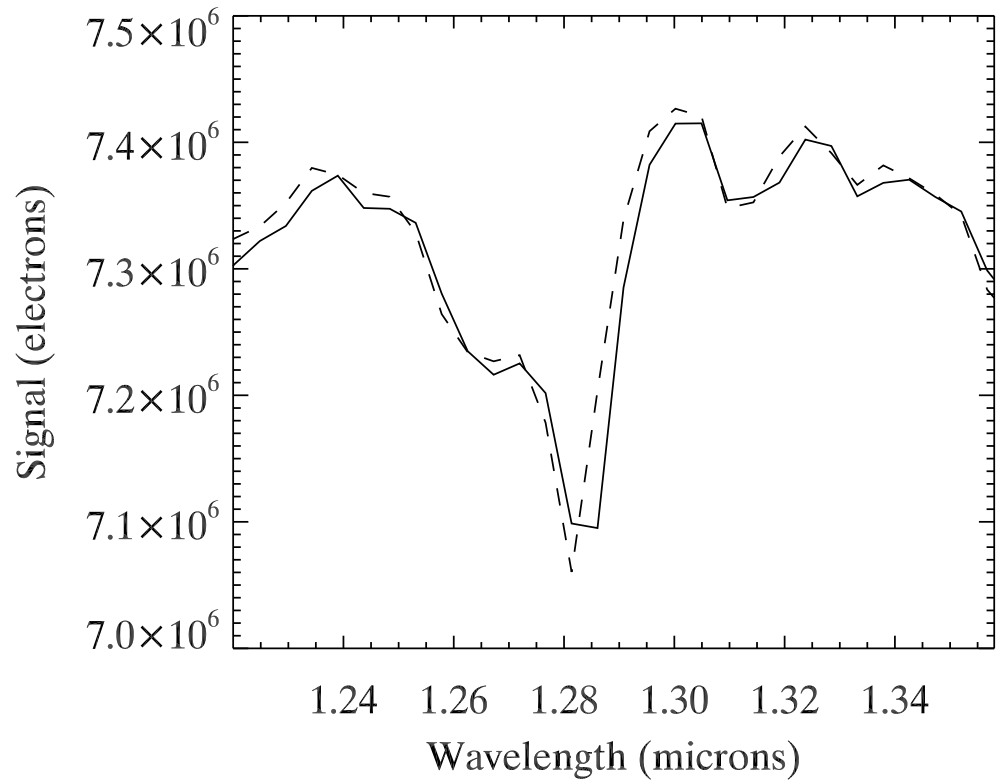}
\caption{Example of undersampling in the WFC3 spectra.  Shown are
zoomed-in portions of two spectra of HD\,209458, separated by 3.1
hours of time (in orbits 1 and 3 of visit 26 in our program).  The dip at
1.28\,$\mu$m is the Paschen-beta line in the star.  In the earlier
spectrum (solid line), the line core appears flattened because the
line is positioned mid-way between two columns of the detector.  In
the later spectrum (dashed line), the line core is sharp because the
different overall shift in wavelength places it centered on a column.
\label{fig6}}
\end{figure}

\clearpage

\begin{figure}
\epsscale{.8}
\hspace{0.5in}
\plottwo{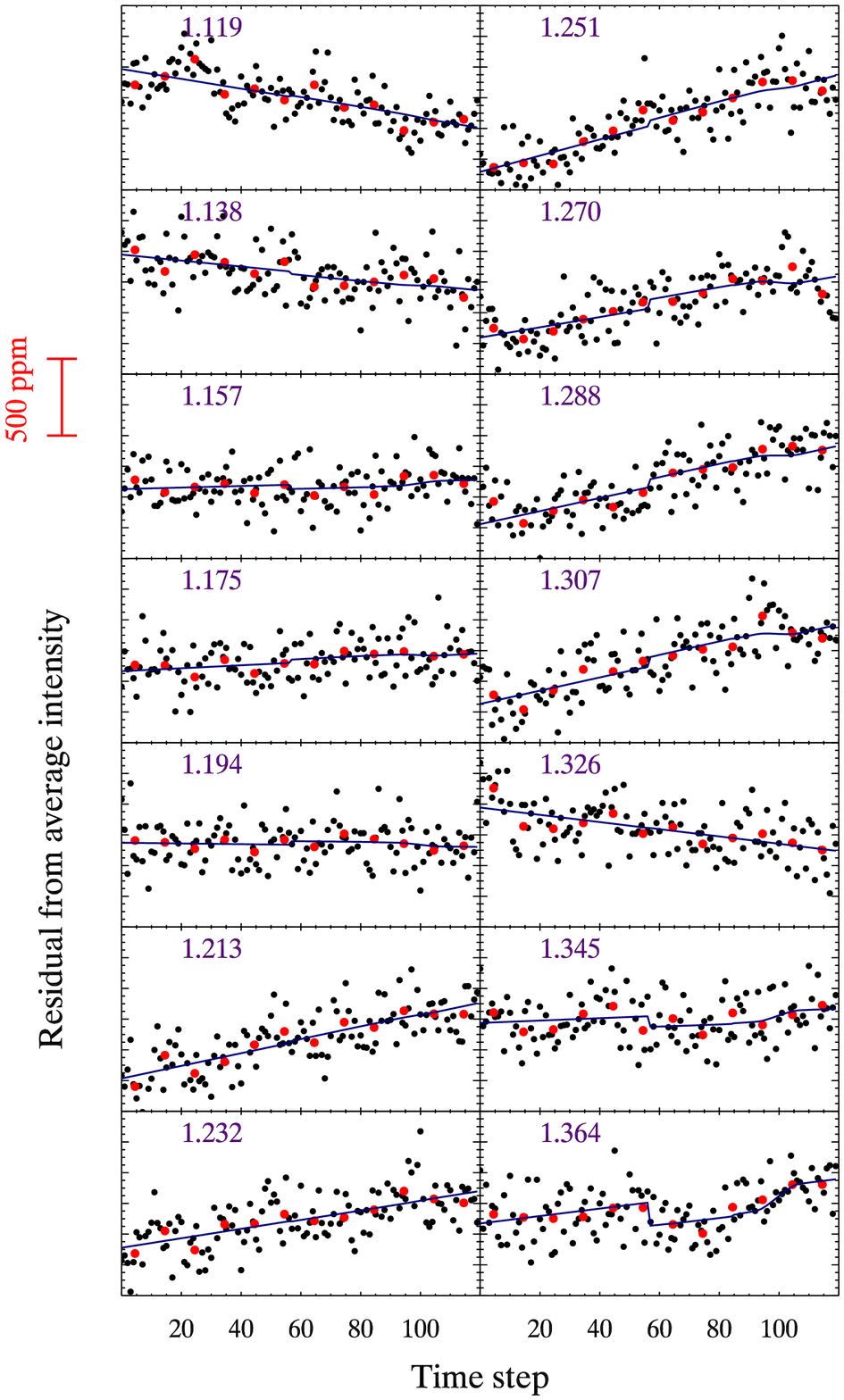}{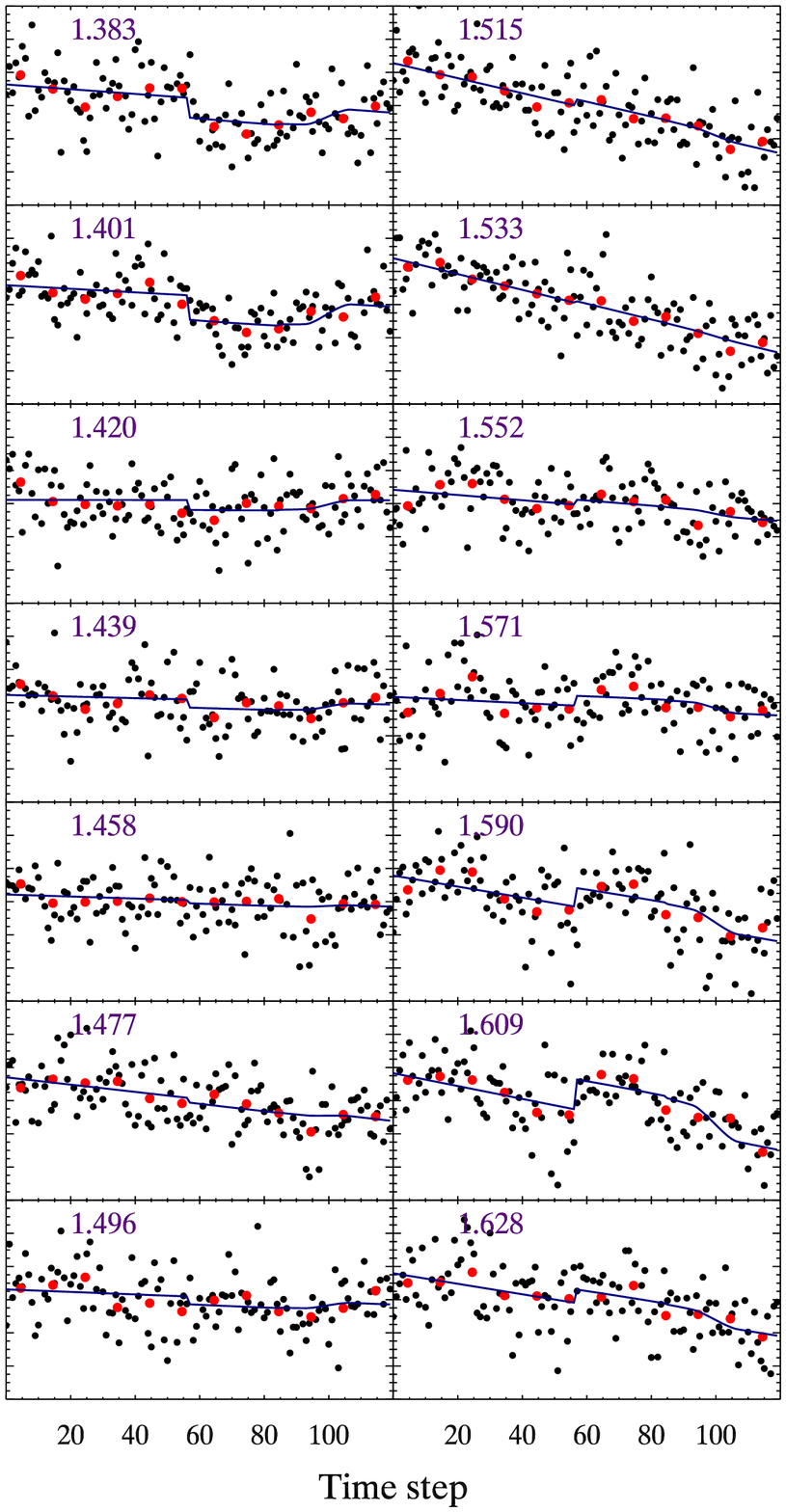}
\vspace{0.7in}
\linespread{1.0}
\caption{Fits of differential transit curves to spectral intensity
residuals for HD\,209458b.  Plotted are values of residuals
$R_{t{\lambda}}$, binned over 4 values of $\lambda$.  In practice, we
fit the differential phase curves to data at {\it individual}
$\lambda$ values, then we bin the fitted amplitudes to form the
transmission spectrum.  These panels show the binned data compared to
a binning of the fits for the 4 wavelengths in each bin.  Each panel
is labeled by binned wavelength in microns.  For all panels, the red
points are temporal averages of 10 data points and are plotted for
illustration purposes only; no temporal averaging is used in the
fitting process.  The fitted curves (in blue) include a linear
baseline as a function of ordinal time step, as well as the
differential transit.  The differential transit can appear distorted
when plotted versus ordinal time step (see Sec.~4.4 discussion of
ordinal {\it vs} phase baselines), but the transit curve is generated
correctly as a function of orbital phase.  Note the obvious increase
in differential transit depth near the bandhead wavelength at
$\sim$\,1.38\,$\mu$m (compare to upper panel of Figure~10).  Also,
note that these differential transit depths are illustrated prior to
the correction for wavelength-dependent limb darkening. Limb
darkening increases transit depths with decreasing wavelength.
\label{fig7}}
\end{figure}
\clearpage

\begin{figure}
\epsscale{.50}
\vspace{-0.5in}
\plotone{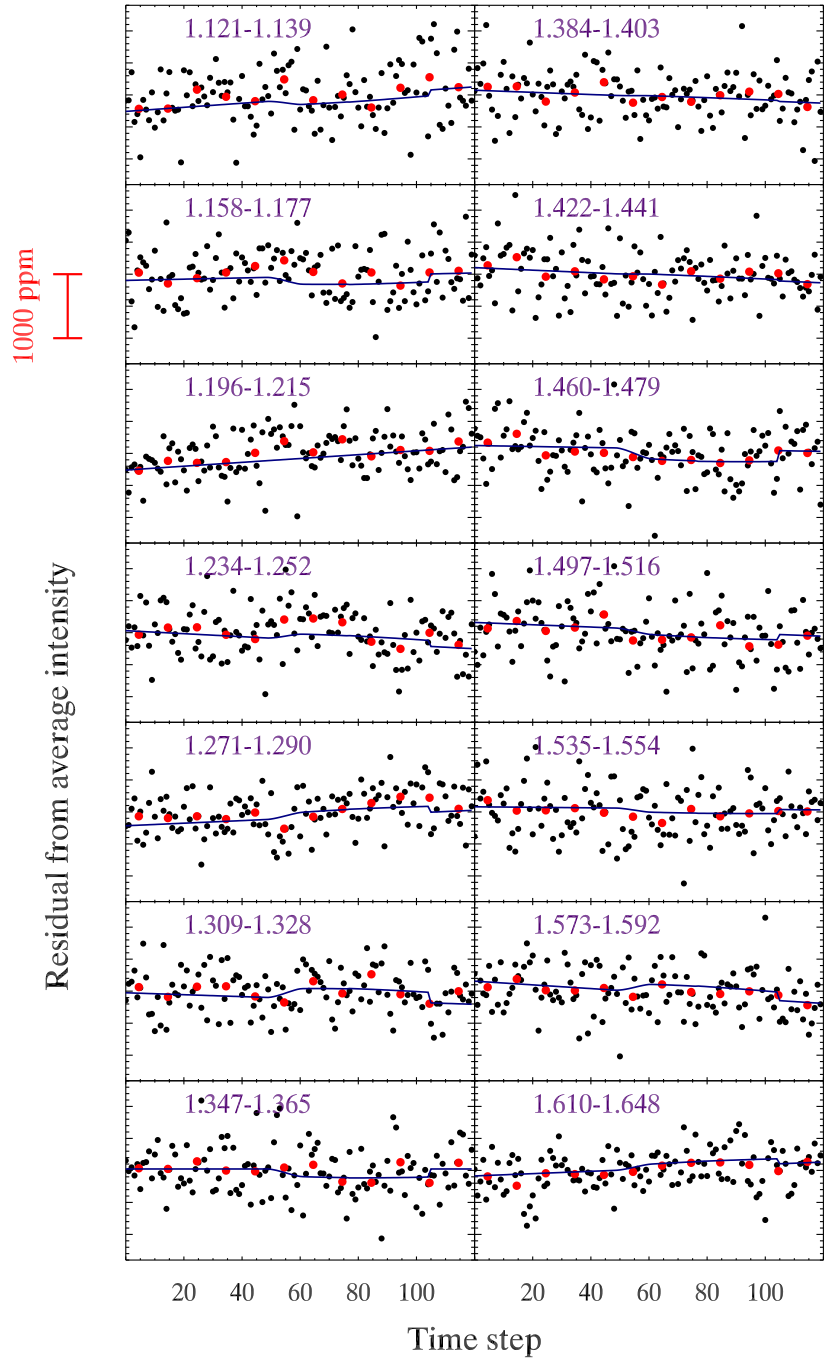}
\vspace{0.7in}
\caption{Fits of differential transit curves to spectral intensity
residuals for XO-1b (see caption of Figure~7 for explanation). To
reduce the scatter, and improve the clarity for this fainter system,
most panels show the average of two wavelength bins (compare to
Table~3), except the lower right panel that includes three wavelength
bins.
\label{fig8}}
\end{figure}

\clearpage

\begin{figure}
\epsscale{.55}
\plotone{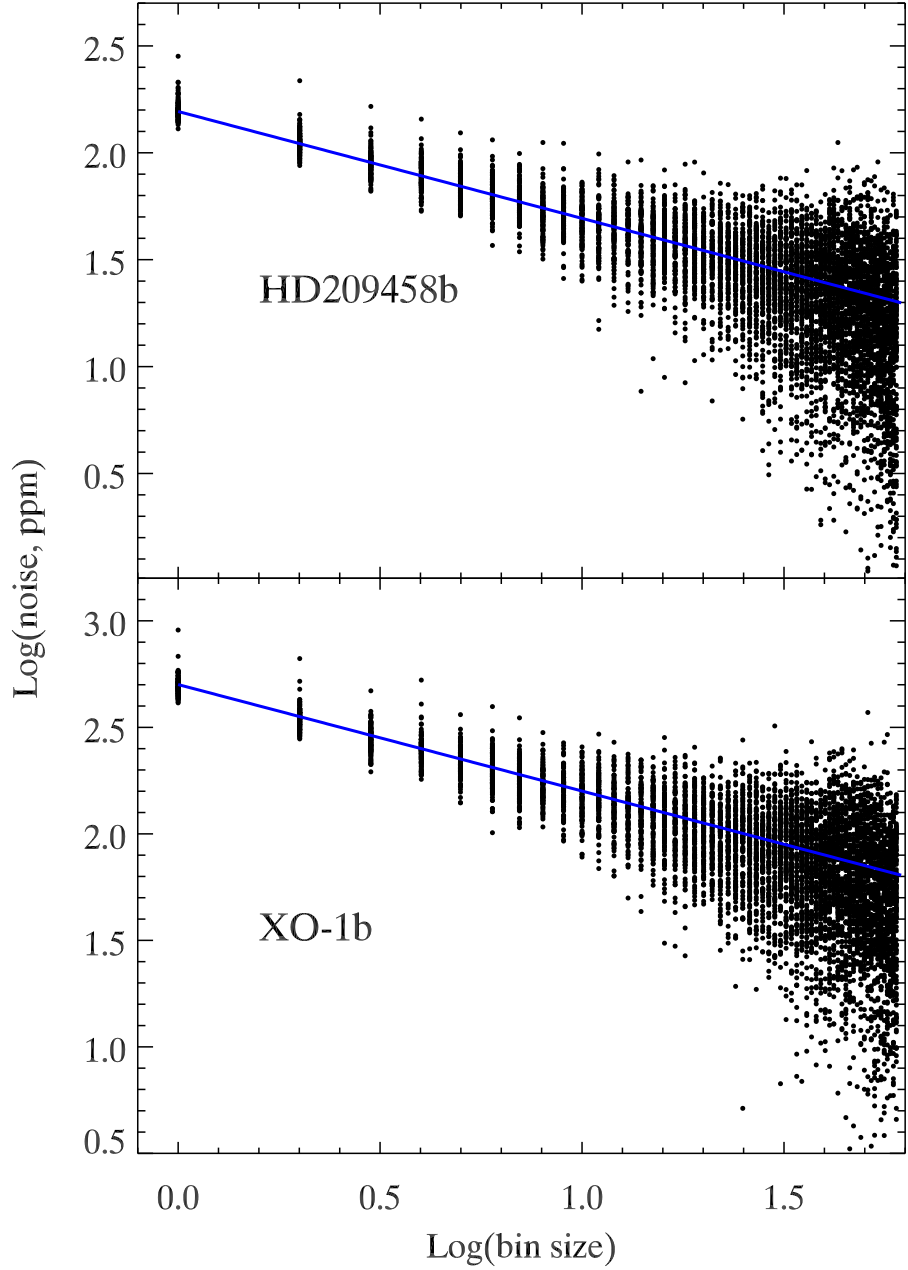}
\vspace{0.7in}
\caption{Error analysis for our derived exoplanetary transmission
spectra.  Each panel plots the standard deviation of the observed
noise in our differential transits, after removing the best-fit
amplitude.  The noise is shown as a function of bin size.  The blue
lines are the relations expected for photon noise based on the number
of detected electrons, and accounting for the effect of smoothing the
grism spectra (Sec.~4.3).  The blue lines have a slope of -0.5 due to
the expected inverse square-root dependence of the noise; the measured
points are in good agreement with that expectation.
\label{fig9}}
\end{figure}

\clearpage

\begin{figure}
\epsscale{.50}
\vspace{-0.5in}
\plotone{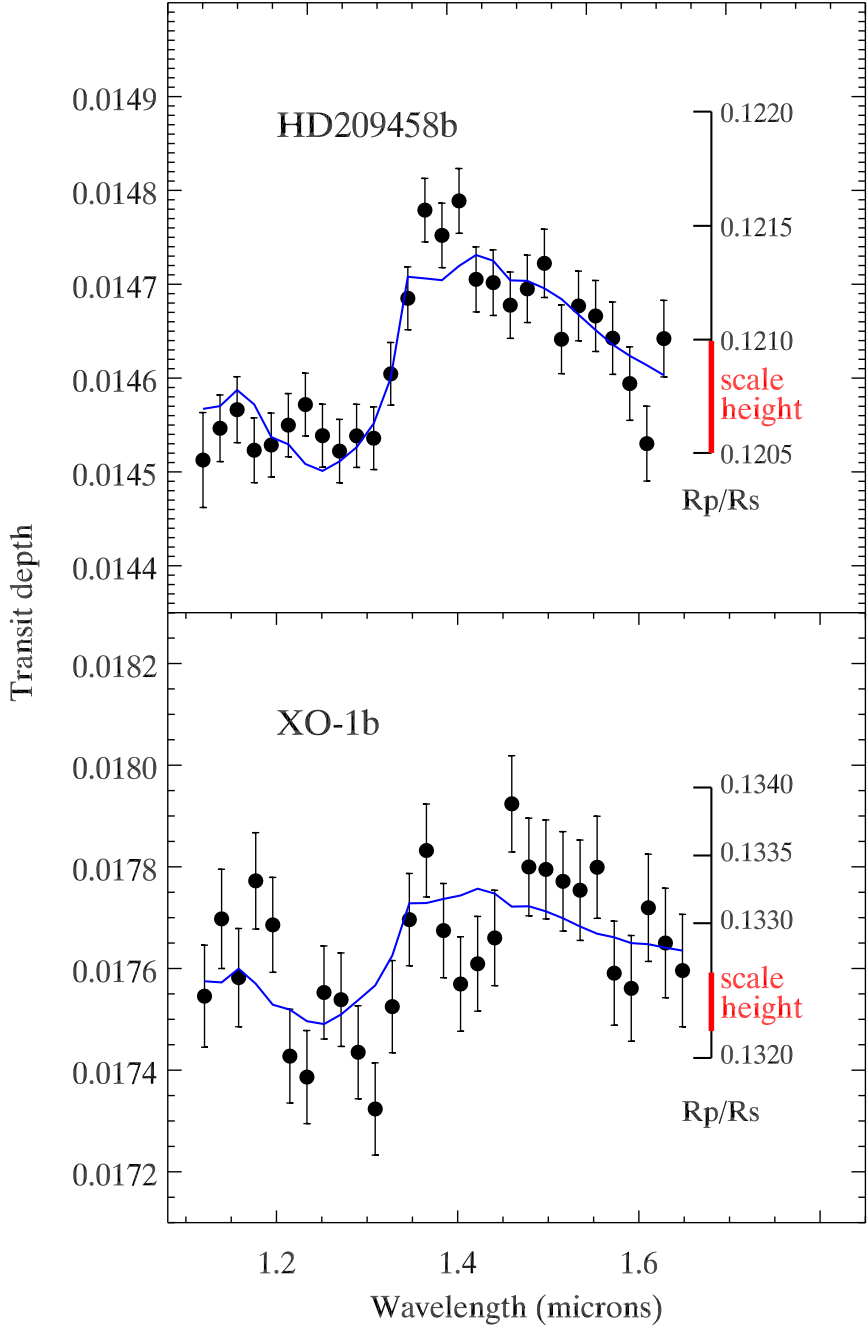}
\vspace{0.7in}
\caption{Our results for transmission spectra for HD\,209458b and
XO-1b in the WFC3 bandpass, compared to models based on Spitzer
secondary observations (blue lines). The spectral resolving power of
these measurements is $\lambda/\delta(\lambda) \approx 70$.  The
amplitude of the 1.4\,$\mu$m water absorption is about 200
parts-per-million (ppm) in both cases, but the errors are smaller for
HD\,209458b due to the greater photon flux.  The ordinate (transit
depth) is $R_p^2/R_s^2$, but $R_p/R_s$ is shown by the scale on the
right, and the red bars indicate the pressure scale heights for both
planetary atmospheres.  The water absorption we detect is about two
pressure scale heights.
\label{fig10}}
\end{figure}

\clearpage

\begin{figure}
\epsscale{.80}
\plotone{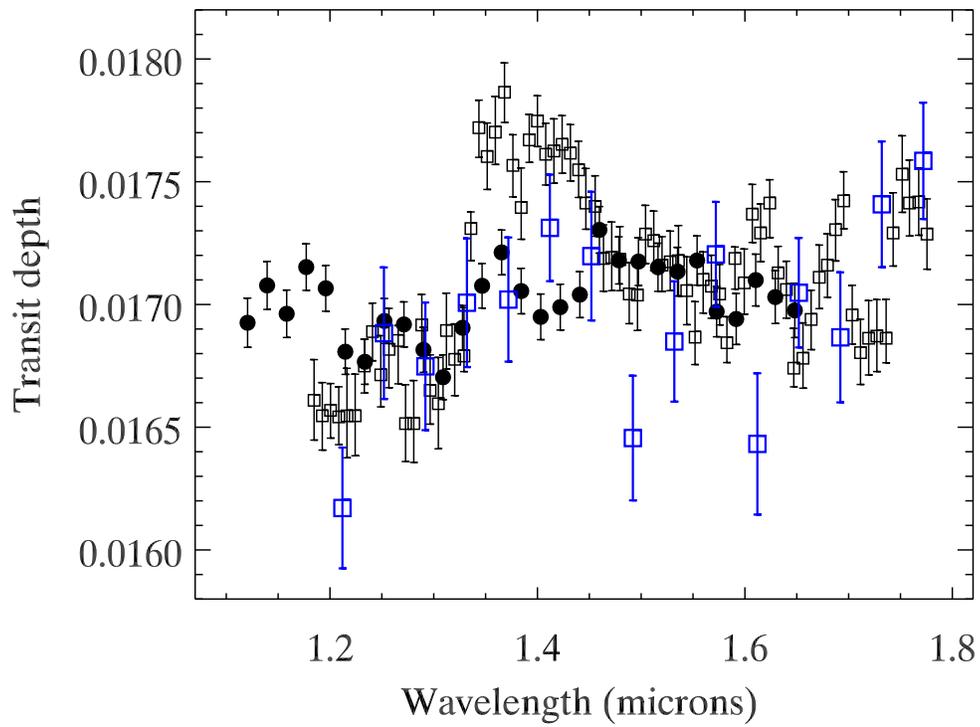}
\caption{Comparison of our transmission spectrum of XO-1b (solid
points, see Figure~10) with the NICMOS results from \citet{tinetti}
(black squares) and \citet{crouzet} (blue squares).  (Our data have
been offset in transit depth for clarity.) Our WFC3 water
absorption is of much smaller amplitude than seen in the NICMOS data
(see text for discussion).
\label{fig11}}
\end{figure}

\clearpage

\begin{figure}
\epsscale{.80}
\plotone{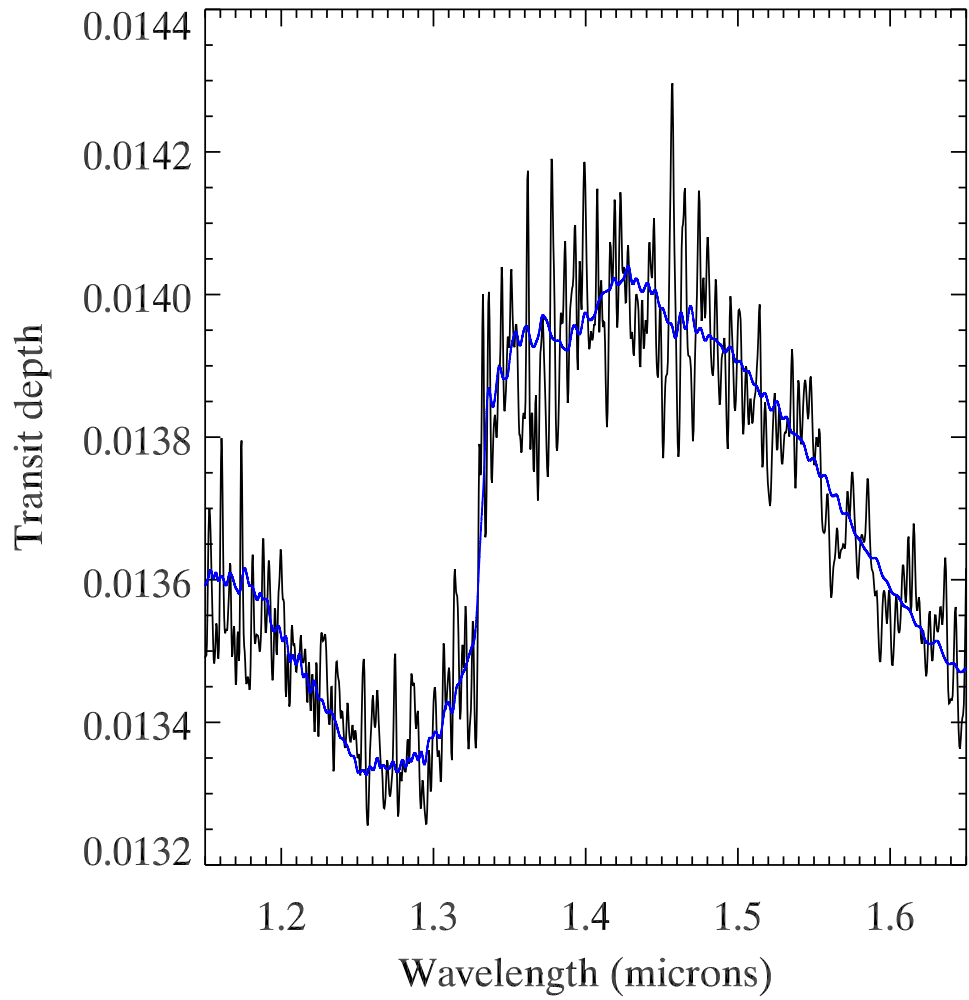}
\caption{Validation of our transmission spectral model (blue line)
versus an independent calculation from \citet{fortney10} for an isothermal model at 1500K.
\label{fig12}}
\end{figure}

\clearpage

\begin{figure}
\epsscale{.80}
\plotone{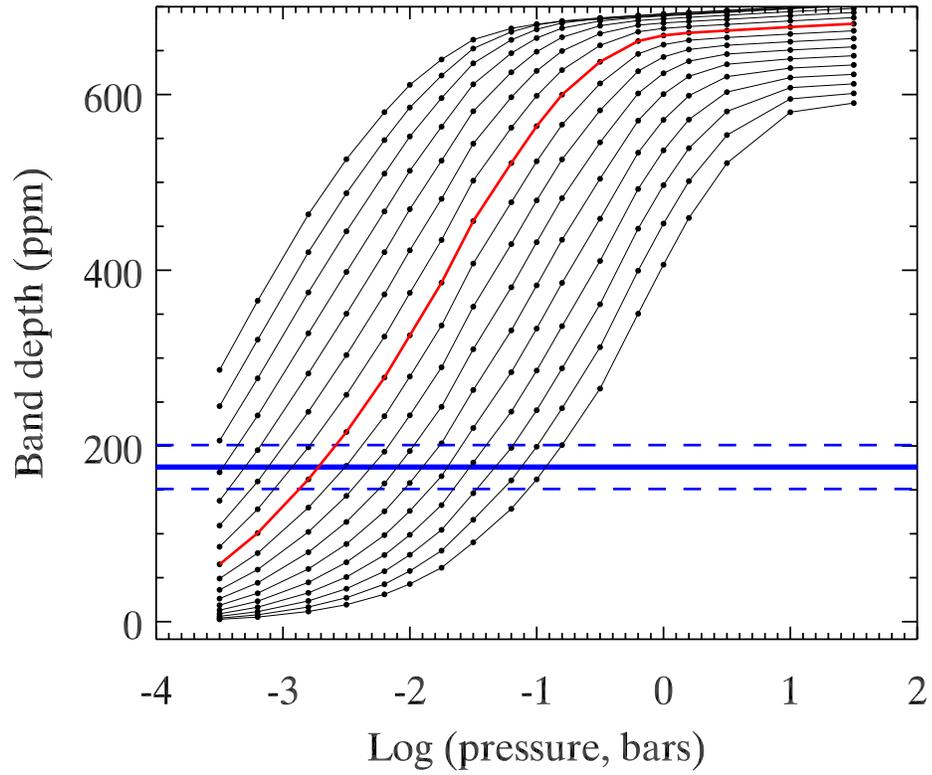}
\caption{Depth of the 1.4\,$\mu$m water band versus cloud top pressure
for an isothermal model at 1200K. Lines represent different mixing
ratios of water. From top to bottom the log of the mixing ratios vary
from -2.0 to -5.2 in increments of -0.2.  The red contour is the
mixing ratio expected for solar abundance. Small black points are the
actual calculations from our transmittance model. The blue line is our
observed band depth of 176 ppm, with $\pm1\sigma$ errors (dashed
lines).  For this Figure, band depth is defined as the average transit
depth from 1.36-1.44\,$\mu$m minus the average depth from
1.27-1.30\,$\mu$m.
\label{fig13}}
\end{figure}

\clearpage

\begin{figure}
\epsscale{.80}
\plotone{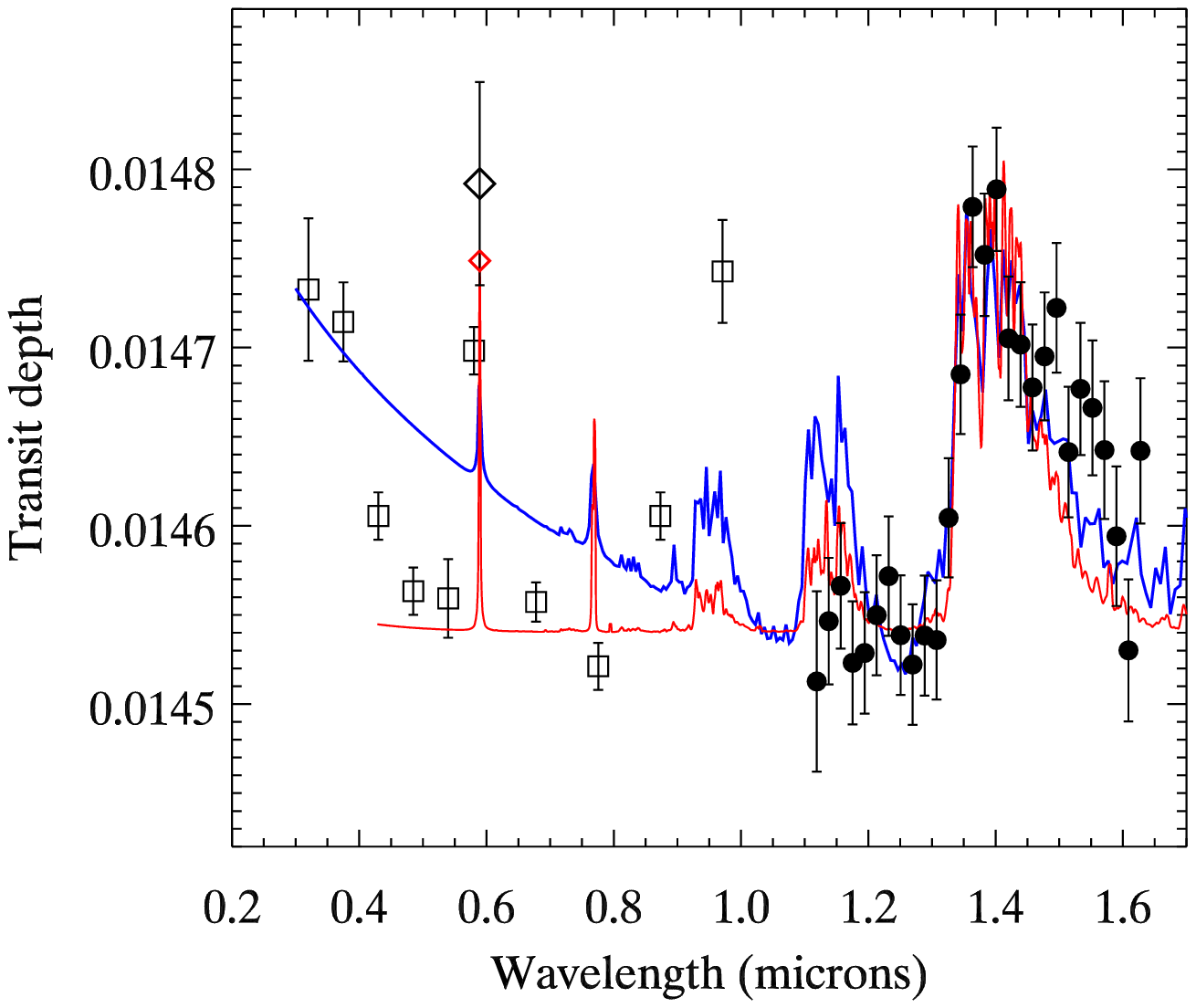}
\caption{Transmission spectrum of HD\,209458b derived from Hubble
spectroscopy.  Our WFC3 results are the solid points. The open squares
are our reanalysis of the STIS bands defined by \citet{knutson07a},
and the diamond is the narrow sodium band absorption from
\citet{charb02}. The red line is the transmittance spectrum from an
isothermal Burrows model, having an extra opacity of gray character
and magnitude 0.012\,cm$^2$\,g$^{-1}$.  The red diamond integrates the
red model over the sodium bandpass.  The blue line is a Dobbs-Dixon
model for HD\,209458b, with no gray opacity, but with $\lambda^{-4}$
(Rayleigh) opacity, normalized to magnitude 0.001\,cm$^2$\,g$^{-1}$ at
0.8\,$\mu$m.  Because the blue model has no gray opacity, we
scale-down the modulation in this spectrum by a factor of 3 for this
comparison (see text).
\label{fig14}}
\end{figure}

\clearpage

\begin{table}
\begin{center}
\caption{Summary of the spatial scan observations \label{tbl-1}}
\begin{tabular}{lcc}
\tableline
\tableline
     & XO-1b & HD\,209458b \\
\tableline
 Time of first scan BJD\,(TDB)  & 2455834.6666 & 2456196.0895 \\
 Planetary orbital phase at first scan  & -0.0471 & -0.0566 \\ 
 Time of last scan BJD\,(TDB)  & 2455834.9419 & 2456196.3997 \\ 
 Planetary orbital phase at last scan  &  0.0228 &  0.0314 \\
 Number of scans  &  128 & 125 \\
 Number of HST orbits & 5 & 5 \\
 Scan rate (arcsec per sec)[pixels per sec] & (0.05)[0.41] & (0.9)[7.44] \\
 Detectory subarray size & 128x128 & 256x256 \\
 Detector reads per scan & 8  & 4 \\
 Duration of scan (sec) & 50.4 & 32.9 \\
 Signal level on detector (electrons per pixel) & $3.8 \times 10^4$ & $4.4 \times 10^4$ \\ 
\tableline
\end{tabular}
\end{center}
Note: planetary orbital phase is defined to be zero at mid-transit.
\end{table}

\begin{table}
\begin{center}
\caption{Results for Radius Ratios ($R_p/R_s$)and Mid-Transit Times \label{tbl-2}}
\begin{tabular}{lcc}
\tableline
\tableline
     & XO-1b & HD\,209458b \\
\tableline
 Mid-Transit Time BJD\,(TDB)  & $2455834.85186\pm0.00017$ & $2456196.28934\pm0.00018$ \\
 $R_p/R_s$  & $0.1328\pm0.0006$ & $0.1209\pm0.0004$ \\ 
\tableline
\end{tabular}
\end{center}
\end{table}

\clearpage

\begin{table}
\begin{center}
\caption{Results for transmission spectra.  Wavelength ($\lambda$) is
 in microns, and transit depth in parts-per-million (ppm).  Note that
 the tabulated errors apply to the differential transit depths; a
 larger error applies to radius ratio over the entire range - see
 Table~2.  (Our re-analyzed STIS transit depths are not listed here,
 but are uniformly 763 ppm larger than given by \citealp{knutson07a}).
\vspace{0.2in}
\label{tbl-3}}
\begin{tabular}{rllrll}
\tableline
\tableline
     & HD\,209458b &   &   & XO-1b &  \\
\tableline
 $\lambda$  & $R_p^2/R_s^2$ (ppm) & Error (ppm) &
   $\lambda$  & $R_p^2/R_s^2$ (ppm) & Error (ppm)  \\ 
\tableline
 1.119 &  14512.7 &  50.6  &  1.121 & 17545.5 &  100.4 \\ [-1ex]
 1.138 &  14546.5 &  35.5  &  1.139 & 17697.6 &  97.6  \\ [-1ex]
 1.157 &  14566.3 &  35.2  &  1.158 & 17582.1 &  96.7 \\ [-1ex]
 1.175 &  14523.1 &  34.6  &  1.177 & 17772.4 &  94.8 \\ [-1ex]
 1.194 &  14528.7 &  34.1  &  1.196 & 17685.8 &  93.4 \\ [-1ex]
 1.213 &  14549.9 &  33.7  &  1.215 & 17427.6 &  92.3 \\ [-1ex]
 1.232 &  14571.8 &  33.5  &  1.234 & 17386.4 &  91.6 \\ [-1ex]
 1.251 &  14538.6 &  33.6  &  1.252 & 17552.8 &  91.6 \\  [-1ex]
 1.270 &  14522.2 &  33.8  &  1.271 & 17538.6 &  92.0 \\  [-1ex]
 1.288 &  14538.4 &  33.7  &  1.290 & 17435.2 &  91.5 \\ [-1ex]
 1.307 &  14535.9 &  33.4  &  1.309 & 17323.6 &  90.8 \\ [-1ex]
 1.326 &  14604.5 &  33.4  &  1.328 & 17525.0 &  90.7 \\ [-1ex]
 1.345 &  14685.0 &  33.5  &  1.347 & 17696.1 &  90.7 \\ [-1ex]
 1.364 &  14779.0 &  33.9  &  1.365 & 17832.1 &  91.4 \\ [-1ex]
 1.383 &  14752.1 &  34.4  &  1.384 & 17674.6 &  92.6 \\ [-1ex]
 1.401 &  14788.8 &  34.5  &  1.403 & 17569.4 &  93.0 \\ [-1ex]
 1.420 &  14705.2 &  34.7  &  1.422 & 17609.2 &  93.2 \\ [-1ex]
 1.439 &  14701.7 &  35.0  &  1.441 & 17660.1 &  93.8 \\ [-1ex]
 1.458 &  14677.7 &  35.4  &  1.460 & 17923.9 &  94.7 \\ [-1ex]
 1.477 &  14695.1 &  35.9  &  1.479 & 17799.7 &  96.1 \\ [-1ex]
 1.496 &  14722.3 &  36.4  &  1.497 & 17794.9 &  97.3 \\ [-1ex]
 1.515 &  14641.4 &  36.6  &  1.516 & 17771.4 &  97.9 \\ [-1ex]
 1.533 &  14676.8 &  37.1  &  1.535 & 17753.9 &  98.7 \\ [-1ex]
 1.552 &  14666.2 &  37.8  &  1.554 & 17799.1 & 100.4 \\ [-1ex]
 1.571 &  14642.5 &  38.6  &  1.573 & 17590.7 & 102.4 \\ [-1ex]
 1.590 &  14594.1 &  39.2  &  1.592 & 17560.9 & 104.0 \\ [-1ex]
 1.609 &  14530.1 &  39.9  &  1.610 & 17719.4 & 105.5 \\ [-1ex]
 1.628 &  14642.1 &  40.8  &  1.629 & 17650.2 & 107.7 \\ [-1ex]
       &          &        &  1.648 & 17595.9 & 110.6 \\ [-1ex]

\end{tabular}
\end{center}
\end{table}

\clearpage






\end{document}